\documentclass[aps, prx, reprint, showpacs, superscriptaddress, longbibliography,a4paper]{revtex4-1}

\usepackage{graphicx}
\usepackage{amsmath, amssymb, braket, mathrsfs, color}
\usepackage[colorlinks,bookmarks=true,citecolor=blue,linkcolor=red,urlcolor=blue]{hyperref}

\newcommand{\beq}{\begin{equation}}
\newcommand{\eeq}{\end{equation}}

\newcommand{\eq}[1]{Eq.~(\ref{#1})}

\newcommand{\tr}{\text{tr}}

\newcommand{\rank}{\text{rank}\,}
\newcommand{\Mat}{\mathrm{Mat}}

\newcommand{\abs}[1]{\left| #1 \right|}
\newcommand{\norm}[1]{\left\Vert #1 \right\Vert}

\newcommand{\vecenv}[2]{\left( \begin{array}{c} #1 \\ #2 \end{array} \right)}
\newcommand{\diag}[1]{\text{diag}\left\{ #1 \right\}}
\newcommand{\spec}[1]{\text{Spec} \! \left[#1\right]}

\newcommand{\wt}[1]{\widetilde{#1}}
\newcommand{\cn}{\colon}
\newcommand{\e}{\mathrm{e}}

\newcommand{\cmplx}{\mathbb{C}}
\newcommand{\intg}{\mathbb{Z}}
\newcommand{\real}{\mathbb{R}}
\newcommand{\torus}{\mathbb{T}}

\newcommand{\id}{\openone}
\newcommand{\nullv}{\mathbf{0}}
\newcommand{\symp}{\mathcal{J}}
\newcommand{\proj}{\mathcal{P}}
\newcommand{\uop}{\mathcal{U}}

\newcommand{\U}{\mathrm{U}}

\newcommand{\hlt}{\mathcal{H}}

\newcommand{\green}{\mathcal{G}}
\newcommand{\ve}{\varepsilon}

\newcommand{\dg}{{\dagger}}
\newcommand{\pdg}{{\phantom{\dagger}}}
\renewcommand{\c}{c^\pdg}
\newcommand{\cd}{c^\dg}
\newcommand{\vc}{\mathbf{c}^\pdg}
\newcommand{\vcd}{\mathbf{c}^\dg}

\newcommand{\viz}{\emph{viz}}

\newcommand{\va}{\mathbf{a}}

\newcommand{\vk}{\mathbf{k}}

\newcommand{\vv}{\mathbf{v}}
\newcommand{\vw}{\mathbf{w}}

\newcommand{\bal}{{\boldsymbol{\alpha}}}
\newcommand{\bbe}{{\boldsymbol{\beta}}}

\newcommand{\bet}{{\boldsymbol{\eta}}}

\newcommand{\bvph}{\boldsymbol{\varphi}}

\newcommand{\bsg}{\boldsymbol{\sigma}}

\newcommand{\ndof}{q}
\newcommand{\nuc}{\mathfrak{n}}

\newcommand{\PT}{\mathrm{PT}}
\newcommand{\PTop}{\mathcal{PT}}
\renewcommand{\L}{\mathrm{L}}
\newcommand{\R}{\mathrm{R}}
\newcommand{\B}{\mathrm{B}}

\newcommand{\nhh}{\mathbf{h}}
\newcommand{\rsurf}{\mathfrak{R}}

\graphicspath{{./}{./figures/}}

\begin{document}
\title{Non-Hermitian systems and topology: A transfer-matrix perspective}

\author{Flore K. Kunst}
\email{flore.kunst@fysik.su.se}
\affiliation{Department of Physics, Stockholm University, AlbaNova University Center, 106 91 Stockholm, Sweden}

\author{Vatsal Dwivedi}
\email{vdwivedi@thp.uni-koeln.de}
\affiliation{Institute for Theoretical Physics, University of Cologne, 50937 Cologne, Germany}

\begin{abstract}
Topological phases of Hermitian systems are known to exhibit intriguing properties such as the presence of robust boundary states and the famed bulk-boundary correspondence. These features can change drastically for their non-Hermitian generalizations, as exemplified by a general breakdown of bulk-boundary correspondence and a localization of all states at the boundary, termed the non-Hermitian skin effect. In this paper, we present a completely analytical unifying framework for studying these systems using generalized transfer matrices -- a real-space approach suitable for systems with periodic as well as open boundary conditions. We show that various qualitative properties of these systems can be easily deduced from the transfer matrix. For instance, the connection between the breakdown of the conventional bulk-boundary correspondence and the existence of a non-Hermitian skin effect, previously observed numerically, is traced back to the transfer matrix having a determinant not equal to unity. The vanishing of this determinant signals real-space exceptional points, whose order scales with the system size. We also derive previously proposed topological invariants such as the biorthogonal polarization and the Chern number computed on a complexified Brillouin zone. Finally, we define an invariant for and thereby clarify the meaning of topologically protected boundary modes for non-Hermitian systems. 

\end{abstract}

\maketitle

\section{Introduction}

A fundamental tenant of quantum mechanics is the reality of the spectra of operators that describe observables, which is typically achieved by demanding these operators to be Hermitian. Abandoning Hermiticity, however, has proved useful in constructing effective descriptions of dissipative systems \cite{rotter_nh_rev, rotter_open_systems_rev}, where non-Hermitian operators encode the interactions with the environment, so that the imaginary part of their spectra can be assigned physical meaning. For instance, the imaginary part of the ``energy'' can be interpreted as the inverse lifetime of a (quasi-)particle \cite{gamow_alpha, landau-lifshitz_qm}. 

The study of non-Hermitian systems has primarily been driven by experiments in photonics \cite{kip_pt_observation_optics, yang_directional_lasing, yang_whispering_gallery,lin_unidir_invis, regensburger_photonic_lattice, feng_single_mode_lasing, hodaei_single_mode_lasing,chen_enhanced_sensing, hodaei_enhanced_sensing, gao_polariton_ep, zhen_ep_ring, weimann_photonic_ssh, zeuner_nh_bulk_transition, schomerus_pt_ssh, poli_sel_enh}, where non-Hermiticity can be realized by judiciously incorporating gain and loss \cite{el_ganainy_non_hermitian, lu_topological_photonics, schomerus_defect_modes}. These setups thus provide concrete realizations of non-Hermitian lattice models, such as a photonic analog of the Su-Schrieffer-Heeger model with topologically protected mid-gap states \cite{weimann_photonic_ssh, zeuner_nh_bulk_transition, schomerus_pt_ssh}. Furthermore, non-Hermitian photonic systems can be engineered to operate at ``exceptional points" at which they exhibit intriguing phenomena such as unidirectional transmission \cite{yang_directional_lasing, yang_whispering_gallery}, one-sided invisibility \cite{lin_unidir_invis, regensburger_photonic_lattice}, single-mode lasing \cite{feng_single_mode_lasing, hodaei_single_mode_lasing}, and enhanced sensitivity to perturbations \cite{chen_enhanced_sensing, hodaei_enhanced_sensing}. Similar realizations of non-Hermitian models are also possible in other experimental setups, such as mechanical \cite{samuel_pt_mech}, acoustic \cite{fleury_pt_acoustic, shi_pt_acoustic}, electronic \cite{kottos_pt_elec}, and ultracold atomic \cite{wunner_pt_cold_atom} systems. 

Theoretically, non-Hermitian Hamiltonians have been used to describe condensed matter systems such as Majorana fermions in topological superconductors \cite{ramon_majorana_abs, avila_majorana_abs}, finite lifetime quasiparticles in heavy-fermion systems \cite{kozii_non_hermitian, yoshida_non_hermitian}, and bosonic superconductors \cite{lieu_topological_symmetry}, as well as to simulate the out-of-equilibrium systems described by a Lindblad master equation \cite{daley_qjump, knight_qjump}. In addition, certain symmetries of non-Hermitian Hamiltonians, such as a parity-time ($\PT$) symmetry \cite{bender_pt_symm_intro, *bender_pt_symm_rev, bender_pt_symm_first, *bender_pt_symm,  mathur_pt_symm} or a more general pseudo-Hermiticity \cite{mostafazadeh_pseudo_hermiticity, *mostafazadeh_pseudo_hermitian_qm}, ensure the reality of its spectrum. These Hamiltonians have garnered significant interest in mathematical physics as an analytic continuation of quantum mechanics to the complex plane \cite{bender_complex_qm}. Recently, a new direction of research has been established by investigating these systems from the perspective of topological phases \cite{alvarez_nh_rev, hughes_nh_ti, yuce_topological_phases,  harter_PT_breaking, yuce_PT_symmetric,  xioing_nh_bulk-bdry, flore_biorth, flore_symmetry_protected, carlstroem_exceptional_links, duan_exceptional_rings,  yao_nh_winding, ueda_nh_ktheory,  fu_nonhermitian_top_band, esaki_edge_states, lee_half_winding, shu_winding_geom, ueda_nh_ci, yao_nh_winding, yao_nh_ci,  bardarson_svd_inv, ueda_nh_tsc}. 
 
Noninteracting topological phases of matter have been of much theoretical \cite{bernevig-hughes_book, shen_book, hasan-kane_ti, hasan-kane_ti_tsc, hughes_HgTe,  kane-mele_qsh, hughes_top_qft} and experimental \cite{molenkamp_qsh, hasan_qsh, hasan_3dti} interest over the last decade. Lacking a local order parameter, these phases are characterized by features that stay unchanged under continuous deformations, such as a quantized bulk topological invariant and the appearance of robust states on their boundaries. A particularly profound result is the \emph{bulk-boundary correspondence}, which establishes a direct link between the bulk invariants and the presence of robust boundary modes. Mathematically, this correspondence can be thought of as a relation between the continuous spectrum and the point spectrum of the system with open boundary conditions. 

Non-Hermitian analogs of topological phases often exhibit drastically different physics from their Hermitian counterparts. An example is the existence of the aforementioned exceptional points (EPs) or more general exceptional structures \cite{carlstroem_exceptional_links, flore_symmetry_protected, duan_exceptional_rings}, where a spectral degeneracy is accompanied by a coalescence of the corresponding eigenstates \cite{kato_book, heiss_EPs}. Another remarkable feature is the possibility of a marked difference between the spectra of systems for periodic and open boundary conditions (hereafter PBC and OBC, respectively), in stark contrast to the Hermitian systems. This is accompanied by a piling up of ``bulk" states at the boundaries for a finite system, a phenomenon termed the \emph{non-Hermitian skin effect} \cite{xioing_nh_bulk-bdry, flore_biorth, yao_nh_winding}. For gapped systems with robust boundary modes, this difference in spectra as well as nature of states for PBC and OBC signify a breakdown of the bulk-boundary correspondence.

The problem of restoring the bulk-boundary correspondence by defining a bulk invariant that can predict the existence of topologically protected boundary modes is rather subtle \cite{ueda_nh_ktheory, fu_nonhermitian_top_band}. Various topological invariants have been defined using a generalization of the conventional Berry connection by replacing the standard inner product with a biorthogonal one \cite{lee_half_winding, esaki_edge_states, shu_winding_geom, fu_nonhermitian_top_band, ueda_nh_ci}, but they often fail to predict the existence of robust boundary modes \cite{ueda_nh_ci, bardarson_svd_inv, yao_nh_winding, yao_nh_ci}. This failure can be traced back to the computation of the topological invariant using the continuous spectrum for a periodic system as opposed to that for a system with open boundaries, a distinction that does not exist for Hermitian systems. Indeed, bulk invariants computed taking this into account for certain specific models \cite{yao_nh_winding, yao_nh_ci, yao_nh_winding, lee_anatomy_of} have been shown to correctly predict the existence of the topological boundary modes. A correct understanding of the bulk-boundary correspondence for non-Hermitian systems thus necessitates an understanding of the eigenstates of a finite system with OBC. 

In this paper, we present just such an approach by constructing generalized transfer matrices \cite{hatsugai_cbs,dh_lee_cbs,tauber-delplace_tm, vd-vc_tm} for quasi one-dimensional non-Hermitian tight-binding models. We show that various qualitative features of these models can be readily gleaned off from the determinant of the transfer matrix without any numerical exact diagonalizations. For a given tight-binding model, we can thus directly answer questions such as: 
\begin{itemize}
    \item Is there a difference between the PBC and OBC spectra? If yes, is it always accompanied by the non-Hermitian skin effect? 
    \item Where do the EPs occur in a finite system? Are they at the same parameter values as those in a periodic system? 
    \item How does one define a bulk topological invariant that predicts the existence of robust boundary modes? 
\end{itemize}
The transfer matrix approach also facilitates analytic computation of the full spectrum and wave functions for arbitrary finite non-Hermitian systems with OBC, quantities which so far have only been accessible by numerical computations. The implementation of the symmetries of the tight-binding model on the transfer matrix provides a new lens to view systems with additional symmetry, which can be used to explain similarities between, for instance, Hermitian and $\PT$-symmetric systems.

More concretely, for the transfer matrix $T$, we show that a necessary and sufficient condition for the equality of the bulk spectra for PBC and OBC is $\abs{\det T} = 1$. We prove that this condition is always satisfied for Hermitian Hamiltonians as well as $\PT$-symmetric Hamiltonians in the $\PT$-unbroken phase, thereby explaining the observed qualitative similarities in their behavior. The ``bulk'' states for OBC are shown to vary as $\abs{\Psi_n} \sim \abs{\det T}^{n/2}$, so that they are localized at the left/right boundary of the system for $\abs{\det T} \lessgtr 1$. Thus, the difference between the PBC and OBC spectra is always accompanied by the non-Hermitian skin effect. Finally, for $\det T \to 0, \infty$, the propagation using the transfer matrix becomes unidirectional, which corresponds to a ``real space EP'' whose order scales with the system size, as previously observed in numerical computations \cite{martinez_non_hermitian}. 

The correct bulk topological invariant can then be computed by using these decaying states for OBC, which corresponds to deforming the Brillouin zone by adding a complex part to the momentum, as derived for particular cases from \emph{ad hoc} methods in Refs.~\cite{yao_nh_winding, yao_nh_ci, yao_nh_winding}. A more geometric picture follows from the algebraic nature of the construction, which is used to construct a Riemann surface associated with the complex energy \cite{hatsugai_cbs}. We show that the deformed Brillouin zone used to compute the bulk invariant is then associated with one set of noncontractible loops, while the other set of noncontractible loops are associated with the boundary modes. We thus get a winding number associated with the boundary modes and thereby clarify the meaning of a ``topologically protected'' boundary mode for a non-Hermitian system, where the notion of a gap may be ill defined. Finally, our formalism extends the real-space biorthogonal polarization \cite{flore_biorth} to more general lattice topologies than those considered in Refs.~\cite{flore_surf_state, flore_biorth, flore_hoti, *flore_hoti_two}. 

\begin{figure}
  \includegraphics[width=0.75\columnwidth]{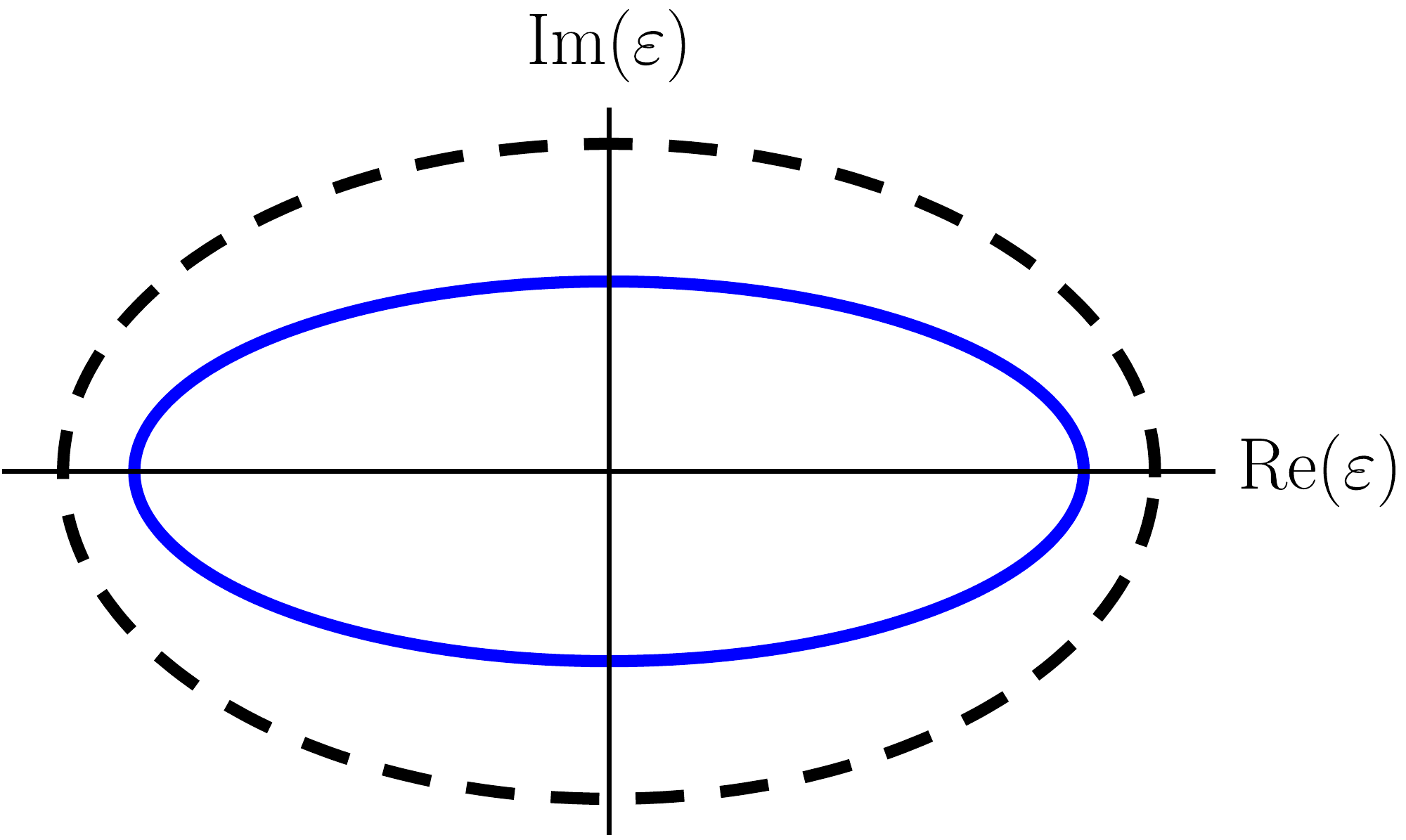}
  \caption{Schematic depiction on the complex $\ve$ plane of the level sets $\abs{\rho(\ve)}=1$ (dashed black curve) and $\abs{\rho(\ve)}=\sqrt{\abs{\det T}}$ (solid blue curve). These curves correspond to the PBC and OBC bulk bands, respectively.}
  \label{fig:schem_broken_BBC}
\end{figure}

Interestingly, the distinction between the PBC and OBC bulk spectra turns out to be quite transparent for cases where the transfer matrix $T(\ve)$ is $2\times 2$ with eigenvalues $\rho_\pm (\ve)$. We show that the PBC bulk bands contain energies $\ve \in \cmplx$ for which $\abs{\rho_a (\ve)} = 1$ for \emph{one of} $a = \pm$, the other naturally satisfying $\abs{\rho_a (\ve)} = \abs{\det T}$. On the other hand, the OBC bulk bands contain energies where $\abs{\rho_a (\ve)} = \sqrt{\abs{\det T}}$ for \emph{both} $a = \pm$. The curves in the complex $\ve$ plane for these two conditions are generically different for $\abs{\det T} \neq 1$, as we depict schematically in Fig.~\ref{fig:schem_broken_BBC}. This explains the difference between the bulk spectra for PBC and OBC for generic non-Hermitian Hamiltonians. On the other hand, we have $\abs{\det T} = 1$ for Hermitian as well as $\PT$-symmetric systems, so that the two curves merge in this case, leading to identical bulk spectra for PBC and OBC. 

The rest of this paper is organized as follows: 
In Sec.~\ref{sec:tm}, we construct the generalized transfer matrix for non-Hermitian tight-binding models and obtain several general results for the spectra. These results are further specialized to a particularly analytically tractable case of $2 \times 2$ transfer matrices in Sec.~\ref{sec:r1}, and an associated energy Riemann surface is constructed. A set of explicit examples illustrating the previously derived general results are presented in Sec.~\ref{sec:eg}. We finally conclude and place this work in a broader context in Sec.~\ref{sec:conc}. Various nonessential details of the calculations are relegated to the appendices. \\

\noindent \emph{Notation:}
We denote the set of $n \times n$ real or complex matrices as $\Mat(n,\real)$ and $\Mat(n,\cmplx)$, respectively. We denote the spectrum of a matrix $M$ by $\spec{M}$.

\section{Transfer matrices}
\label{sec:tm}

Transfer matrices arise naturally in discrete calculus as a representation of recursion relations. Since tight-binding models are essentially composed of hopping, i.e., shift, operators acting on the wave functions, the Schr\"odinger equation for a one-dimensional system can be reduced to a set of recursion relations, which can then be recast into a transfer-matrix equation \cite{hatsugai_cbs, vd-vc_tm}. Thus, starting with a $d$-dimensional system, we impose PBC along $(d-1)$ directions and OBC along the remaining direction, along which the transfer matrix is computed. Choosing the direction of OBC, we can analytically explore the boundary states for various possible boundaries.

\subsection{General setup}  \label{sec:tm_setup}
Consider a system in $d$ spatial dimensions with OBC along $x$ and PBC along the remaining $(d-1)$ directions, which are parametrized by the transverse quasi-momentum $\vk_\perp \in \torus^{d-1}$. This system can also be interpreted as a family of one-dimensional chains parametrized by $\vk_\perp$. Explicitly, we consider a system described by a general tight-binding \emph{non-Hermitian} Hamiltonian
\begin{align}
  \hlt &= \sum_{n=0}^{N_0-1} \sum_{\alpha, \beta=1}^\ndof  \left[ \sum_{\ell=1}^{R} \left( \cd_{n, \alpha} [\mathbf{t}^\pdg_{\L,\ell}]_{\alpha\beta} \c_{n+\ell, \beta} \right. \right. \nonumber \\ 
  & \qquad \left. \left. + \cd_{n+\ell, \alpha} [\mathbf{t}^\dg_{\R,\ell}]_{\alpha\beta} \c_{n, \beta} \right) 
   + \cd_{n, \alpha} [\mathbf{t}^\pdg_{0}]_{\alpha\beta} \c_{n, \beta} \right].
\end{align}
Here, the matrices $\mathbf{t}_{\L,\ell}$ ($\mathbf{t}_{\R,\ell}$) denote the hopping to the left (right) and $\mathbf{t}_0$ is the intra-unit-cell term. For Hermitian systems, these matrices satisfy $\mathbf{t}_{\L,\ell} = \mathbf{t}_{\R,\ell}$ and $\mathbf{t}_0^\dg = \mathbf{t}_0$. The hopping depends only on the distance owing to translation invariance and $R < \infty$ is the range of hopping. We have $\ndof$ internal degrees of freedom, e.g., spin, orbital, or sublattice, per unit cell. The explicit dependence on $\vk_\perp$ is suppressed to avoid notational clutter, however, all parameters should be assumed to depend on $\vk_\perp$, unless stated otherwise.

We reduce this Hamiltonian to a nearest-neighbor form \cite{dh_lee_cbs} by bundling together $\nuc \geq \ndof R$ degrees of freedoms into a \emph{supercell}, whose creation (annihilation) operators are denoted by $\vcd$ ($\vc \!\!$). This definition is not unique, and one may indeed choose arbitrarily large supercells with nearest-neighbor hopping. The Hamiltonian reduces to
\begin{align}
  \hlt(\vk_\perp) &= \sum_{n=0}^{N} \left[ \vcd_{n} J^\pdg_\L  \vc_{n+1} + \vcd_{n} M  \vc_{n} + \vcd_{n+1} J^\dg_\R  \vc_{n} \right]   \label{eq:hlt_gen}
\end{align}
with the \emph{hopping matrices} $J_{\L, \R}$ and the \emph{on-site matrix} $M$, where the latter encodes the hopping between degrees of freedom within the supercell as well as the on-site energies. An arbitrary single-particle state is given by
\begin{equation}
    \ket{\Psi} = \sum_{n=0}^N \Psi_n \vcd_n \ket{\Omega}, \quad   \label{eq:wf_def}
\end{equation}
with $\ket{\Omega}$ the fermionic vacuum state and $\Psi_n \in \cmplx^\nuc$ the wave function for each supercell. The single-particle Schr\"odinger equation $\hlt \ket{\Psi} = \ve \ket{\Psi}$ thus reduces to the recursion relation
\begin{equation}
  J^\pdg_\L \Psi_{n+1} + M \Psi_n + J^\dg_\R \Psi_{n-1} = \ve \Psi_n.
  \label{eq:recur_mat} 
\end{equation} 
We seek to express this as a transfer-matrix equation for cases where $J_{\L,\R}$ may be singular. 

In this paper, we take $M$ to be arbitrary, possibly non-Hermitian, while we demand that the hopping matrices satisfy 
\begin{equation}
    J_\R = J_\L = J, \qquad 
    J^2 = 0,
\end{equation}
For a Hermitian system, $M^\dagger = M$ and $J_\R = J_\L$, so that we have lifted the Hermiticity condition on the on-site matrix $M$ but not the hopping matrix $J$. 
The nilpotence of $J$ implies that no sublattice site within the supercell has hoppings to both left and right adjacent supercells. This can always be ensured by choosing a large enough supercell (also see Ref.~\cite[Appendix~B]{vd-vc_tm}). Mathematically, we need this condition to ensure that the singular vectors of $J$ [Eq.~(\ref{eq:svec})] form an orthonormal set.  Under these assumptions, the recursion relation becomes 
\begin{equation}
 J \Psi_{n+1} + M \Psi_n + J^\dg \Psi_{n-1} = \ve \Psi_n,    \label{eq:recur}
\end{equation}
which corresponds to the (full) Bloch Hamiltonian  
\begin{equation}
    \hlt_\B(\vk) = J(\vk_\perp) \, \e^{ik_x} + M(\vk_\perp) + J^\dg(\vk_\perp) \, \e^{-ik_x}.
    \label{eq:hlt_bloch}
\end{equation}
In practice, we simply use this equation to identify $M$ and $J$ as the coefficients of $\e^{ik_x}$ and 1, respectively, to compute the transfer matrix for propagation along $x$.

\subsection{Constructing the transfer matrix}   \label{sec:tm_const}
We construct the generalized transfer-matrix representation of the recursion relation in Eq.~(\ref{eq:recur}) following Ref.~\cite[Sec.~II]{vd-vc_tm}, which we briefly describe here. The recursion relation can be rewritten as 
\begin{equation}
  \Psi_n = \green J \Psi_{n+1} + \green J^\dg \Psi_{n-1},    \label{eq:recur1}
\end{equation}
where $\green = (\ve\id - M)^{-1}$ is the on-site Green's function, which is nonsingular except when $\ve$ is an eigenvalue of $M$. Next, we compute a reduced singular value decomposition (SVD) \cite[Sec.~6.3]{strang_book}
\begin{equation}
  J = V \, \Xi \, W^\dg, 
\end{equation}
where $\Xi = \text{diag}\{\xi_1, \dots, \xi_r\}$ with $r = \rank{J}$ and the singular values $\xi_i$ are real and positive. Physically, this signifies that a suitable unitary transform of the Hamiltonian reduces it to a form where the consecutive supercells have exactly $r$ hoppings, with the magnitude of the corresponding hopping strengths given by the singular values $\xi_i$'s. 
The $r$ corresponding left and right singular vectors are assembled in the $\nuc \times r$ matrices $V$ and $W$, which satisfy
\begin{equation}\label{eq:svec}
    V^\dg V = W^\dg W = \id_r, \;\; V^\dg W = 0,
\end{equation}
where the orthogonality of $V$ and $W$ follows from $J^2 = 0$, which also ensures that $r \leq \nuc/2$. 

As the vectors in $V$ and $W$ form an orthonormal set, they can be extended 
\footnote{
    We refer to  Ref.~\cite[Sec.~II.B]{vd-vc_tm} for details. Also note that this breaks down if $J_\L \neq J_\R$. 
} 
to an orthonormal basis of $\cmplx^\nuc \ni \Psi_n$. We then define the coefficients of $\Psi_n$ in this basis as
\begin{equation}
    \bal_n = V^\dg \Psi_n, \qquad 
    \bbe_n = W^\dg \Psi_n,  \label{eq:dofs}
\end{equation}
in terms of which Eq.~(\ref{eq:recur1}) becomes
\begin{align}
 \Psi_n  = \green \, V \, \Xi \, \bbe_{n+1} + \green \, W \, \Xi \, \bal_{n-1}.       \label{eq:recur2}
\end{align}
Multiplying to the left by $V^\dg$ and $W^\dg$, we find
\begin{align}
  \bal_n &= \green_{vv} \, \Xi \, \bbe_{n+1} + \green_{wv} \, \Xi \, \bal_{n-1}, \nonumber \\ 
  \bbe_n &= \green_{vw} \, \Xi \, \bbe_{n+1} + \green_{ww} \, \Xi \, \bal_{n-1},
\end{align}
where we have defined $\green_{AB} = B^\dg \, \green \, A \in \Mat(r, \cmplx)$ with $A,B \in \{V, W\}$. This system of equations can be rewritten as 
\begin{equation}
    \Phi_{n+1} = T \Phi_n, \qquad \Phi_n \equiv \vecenv{\bbe_{n}}{\bal_{n-1}},     \label{eq:tm_act}
\end{equation}
where the $2r$-dimensional transfer matrix is given by
\begin{equation} 
   T = 
  \begin{pmatrix}
    \Xi^{-1} \cdot \green_{vw}^{-1} \quad & - \Xi^{-1}  \cdot \green_{vw}^{-1} \cdot \green_{ww} \cdot \Xi \\ 
    \green_{vv} \cdot \green_{vw}^{-1} \quad & \left( \green_{wv} - \green_{vv} \cdot \green_{vw}^{-1} \cdot \green_{ww} \right) \cdot \Xi 
  \end{pmatrix}.
  \label{eq:tm_def}
\end{equation}
The rank of $J$, and hence the size of the transfer matrix is independent of the choice of a supercell \cite[Appendix~B]{vd-vc_tm}. 

Given $\Phi_0$, we can propagate it with the transfer matrix $T$ as
\begin{equation}
    \Phi_n = T^n \Phi_0, \qquad \forall n \in \intg,
\end{equation}
provided $T$ is invertible, i.e., $\det T \neq 0$. 
We explicitly compute 
\begin{equation}
   \det T = \det(\green_{vw}^{-1} \green_{wv}) 
   = \frac{\det \green_{wv}}{\det \green_{vw}}.  \label{eq:dett}
\end{equation}
A distinct possibility for non-Hermitian systems is $\abs{\det T} \to 0, \infty$ when $\abs{\det \green_{wv}} \to 0$ and $\abs{\det \green_{vw}} \to 0$, respectively. Note that these two cases are dual to each other, since if $\abs{\det T } \to \infty$ for some parameters, we can compute the transfer matrix for translation in the opposite direction, whose determinant would then tend to zero. Physically, this corresponds to unidirectionality in the system, since it means that the states can be propagated only in one direction.

The construction above computes the transfer matrix for a right eigenstate. We can perform a similar construction of a transfer matrix for the left eigenstates by considering the action of $\hlt$ on an arbitrary single particle bra state $\bra{\Psi}$, instead of the ket in Eq.~(\ref{eq:wf_def}). Alternatively, we note that the left eigenvectors of $\hlt$ are related to the right eigenvectors of $\hlt^\dg$ by a conjugate transpose. Thus, we can repeat the computation above with a new Bloch Hamiltonian
\begin{equation}
    \wt{\hlt} = \hlt^\dg \implies 
    \wt{\green}(\ve) = \green^\dag(\ve^\ast)
    \label{eq:tm_left_eig}
\end{equation}
to get the transfer matrix for the left eigenstates of $\hlt$.

\subsection{Special cases}     \label{sec:tm_spcl}
The transfer matrix possesses additional structure if the original Hamiltonian is Hermitian or $\PT$-symmetric, as we now show.

\subsubsection{Hermitian systems}
For Hermitian systems, the Bloch Hamiltonian satisfies $\hlt_\B^\dg(\vk) = \hlt_\B^\pdg(\vk)$. For the Bloch Hamiltonian defined in \eq{eq:hlt_bloch}, this implies that $M^\dg = M$ with no additional condition on $J$. We compute $\green^\dg(\ve) \equiv \left[ \green(\ve^\ast) \right]^\dg$ as
\[
    \green^\dg(\ve) = \left[ (\ve^\ast \id - M)^{-1} \right]^\dg =  \left(\ve \id - M^\dg \right)^{-1} = \green(\ve),
\] 
so that $\green_{AB}^\dg(\ve^\ast) = \green_{BA}^\pdg(\ve)$ and Eq.~(\ref{eq:dett}) reduces to
\begin{equation}
	\det T 
	= \frac{\det \green^\pdg_{wv}(\ve)}{\det \green^\dg_{wv}(\ve^\ast)}
	= \frac{\det \green^\pdg_{wv}(\ve)}{ \left[\det \green_{wv}(\ve^\ast) \right]^\ast}. 
\end{equation}
Thus, for $\ve\in\real$, i.e., the regime of physically relevant energies for Hermitian systems, $\det T = \exp[2i\arg \green_{wv}(\ve)]$ lies on the unit circle. As expected, this reproduces Ref.~\cite[Eq.~(26)]{vd-vc_tm}.

\subsubsection{\texorpdfstring{$\PT$}{PT}-symmetric systems} 
\label{sec:ptsymmsystems}
$\PT$-symmetry is implemented as $\PTop = \uop K$ with $\;\uop \in \U(\nuc)$ and $K$ the complex conjugation, so that a $\PT$-symmetric system satisfies $\uop \, \hlt_\B^\ast(\vk) \, \uop^\dg = \hlt_\B^\pdg(\vk)$. Imposing this on the Bloch Hamiltonian in \eq{eq:hlt_bloch}, we find 
\begin{equation}
    J = \uop \, J^T \, \uop^\dg, \qquad 
    M = \uop \, M^\ast \, \uop^\dg.
\end{equation}
Using the condition on the on-site matrix, we can compute $\green^\ast(\ve) \equiv  \left[ \green(\ve^\ast) \right]^\ast$ as 
\[
    \green^\ast(\ve) 
    =  \left(\ve \id - \uop^\dg M \uop \right)^{-1} 
    = \uop^\dg \, \green(\ve) \, \uop.
\] 
We next derive a condition on the singular vectors $V$ and $W$ that satisfy the condition on $J$. We here need to distinguish the two cases corresponding to $\left(\PT\right)^2 = \pm 1$, which are discussed in Appendix~\ref{app:pt_symm}. 

\paragraph{$\left(\PT\right)^2 = +1$:}
In this case, $\uop = \uop^T$ and in Appendix \ref{app:svd}, we show that $V, W$ must satisfy
\[ 
    V = \uop W^\ast, \qquad 
    W = \uop V^\ast,
\] 
which is consistent, since $\uop \uop^\ast = \uop \uop^\dg = \id$. Furthermore, 
\[ 
    J = V \Xi W^\dg = \uop W^\ast \Xi V^T \uop^\dg = \uop J^T \uop^\dg
\] 
as desired. We can now compute 
\begin{align*}
    \green_{vw}^\ast(\ve) 
    &= W^T \green^\ast(\ve) V^\ast  \\ 
    &= V^\dg \uop^T \uop^\dg \green(\ve) \uop \uop^\ast W
    = \green_{wv}(\ve),
\end{align*} 
so that Eq.~(\ref{eq:dett}) reduces to
\begin{equation}
	\det T 
	= \frac{\det \green^\pdg_{wv}(\ve)}{\det \green^\ast_{wv}(\ve)}
	= \frac{\det \green^\pdg_{wv}(\ve)}{\left[ \det \green_{wv}(\ve^\ast) \right]^\ast}, 
\end{equation}
which, as in the Hermitian case, lies on the unit circle for $\ve \in \real$, i.e., in the $\PT$-unbroken phase.

\paragraph{$\left(\PT\right)^2 = -1$:}
In this case, $\uop^T = -\uop$ is even dimensional, as shown in Appendix~\ref{app:pt_symm}. Alternatively, this must be the case since $\uop \in \uop(\nuc) \implies \abs{\det \uop} = 1$, while the determinant vanishes for any odd-dimensional antisymmetric matrix. As we show in Appendix~\ref{app:svd}, the singular values of $J$ must also come in doubly degenerate pairs, so that $\rank J$, i.e., the number of nonzero singular values of $J$, is even, and we can write 
\begin{equation}
    \Xi = \diag{\xi_1 \id_2, \xi_2 \id_2, \ldots, \xi_{r/2} \id_2}.
\end{equation}
We now define 
\begin{equation}
    \Sigma \equiv \diag{\mathscr{J}, \ldots, \mathscr{J}}, \qquad  
    \mathscr{J} = 
    \begin{pmatrix}
        0 & 1 \\ 
        -1 & 0 
    \end{pmatrix}. 
\end{equation}
Here, $\Sigma$ is antisymmetric and satisfies $\Sigma^2 = -\id$ and $[\Sigma, \Xi] = 0$, the latter being the case because $\Xi$ is proportional to the identity matrix in each $2\times 2$ block. In Appendix \ref{app:svd}, we show that $V, W$ satisfy 
\begin{equation*}
    V = \uop W^\ast \Sigma, \qquad 
    W = \uop V^\ast \Sigma,
\end{equation*}
which is consistent since 
\begin{align*}
    V &= \uop (\uop V^\ast \Sigma)^\ast \Sigma 
      = -V \Sigma^2 = V, \\ 
    J &= V \Xi W^\dg       = - \uop W^\ast \Sigma^2 \Xi V^T \uop^\dg  
      = \uop J^T \uop^\dg. 
\end{align*}
Finally, we can compute 
\begin{align*}
    \green_{vw}^\ast(\ve) 
    &= W^T \green^\ast(\ve) V^\ast \\
    &= \Sigma^T V^\dg \uop^T \cdot \uop^\dg \green(\ve) \uop \cdot \uop^\ast W \Sigma \\
    &= -\Sigma \, \green_{wv}(\ve) \, \Sigma.
\end{align*}
Thus, 
\begin{equation}
	\det T 
	= \frac{\det \green^\pdg_{wv}(\ve)}{\det \left[ -\Sigma \, \green_{wv}^\ast(\ve) \, \Sigma \right]}
	= \frac{\det \green^\pdg_{wv}(\ve)}{\left[ \det \green_{wv}(\ve^\ast) \right]^\ast}, 
\end{equation}
since $\det\left[ -\Sigma^2 \right] = \det \id = 1$. As in Hermitian case, $\det T$ lies on the unit circle for $\ve \in \real$. 

In conclusion, the presence of either Hermiticity or a $\PT$-symmetry implies the unimodularity of the transfer matrix. This is the precise sense in which the two systems behave in a similar fashion. Other symmetries of non-Hermitian Hamiltonians may also lead to this similarity with Hermitian systems, e.g., for parity-particle-hole (CP) symmetry which takes $\hlt_\B (\vk) \to - \uop \, \hlt_\B^\ast(\vk) \, \uop^\dg$, we find
\begin{equation*}
	\det T 
	= \frac{\det \green^\pdg_{wv}(\ve)}{\left[ \det \green_{wv}(-\ve^\ast) \right]^\ast}, 
\end{equation*}
so that $T$ is unimodular if $\ve \in i \real$.

\subsection{Spectra and states}    \label{sec:tm_spec}
The spectrum of the transfer matrix for a given $(\ve, \vk_\perp)$ contains information about the possible states for that specific energy $\ve$. This can also be thought of as a discrete scattering problem, where for an incoming ``plane wave'' of a given energy, the spectrum of the transfer matrix contains information about the fate of that plane wave as it propagates through the system. The eigenstates of the systems can then be thought of as the stationary or \emph{standing-wave} solutions. Given a boundary condition, the task then is to find the values $(\ve, \vk_\perp)$ that are compatible with such stationary solutions. 

For condensed-matter systems, the most common boundary conditions to consider are periodic (PBC) and open (OBC) ones. In the following, we start with a ring with $N$ supercells realizing PBC and consider an interpolation between these two cases by tuning the strength of one of the bonds continuously to zero.

\subsubsection{Periodic boundary condition}
For a periodic system with $N$ supercells, $\Psi_n = \Psi_{n+N}$, so that using Eq.~(\ref{eq:tm_act}), we must have
\begin{equation}
    \Phi_{n} = \Phi_{n+N} \implies \Phi_n = T^N(\ve, \vk_\perp) \Phi_n.
\end{equation}
Thus, the system with PBC has a state for a given $(\ve, \vk_\perp)$ \emph{iff} $1 \in \spec{T^N(\ve, \vk_\perp)}$, which reduces to 
\begin{equation}
     \e^{2\pi i \ell/N} \in \spec{T(\ve,\vk_\perp)}
    \label{eq:pbc1}
\end{equation}
for some $\ell \in \{0, \dots, N-1\}$. As $N\to\infty$, the set of these points is dense on the unit circle. Thus, the bulk band for a given $\vk_\perp$ is the closed, compact set of $\cmplx \ni \ve$ for which at least one eigenvalue $\rho$ of $T(\ve,\vk_\perp)$ lies on the unit circle. Setting $\rho = \e^{ik_x}$ and $\Phi_0 = \bvph$ as the corresponding eigenvector (or one of the eigenvectors, if the corresponding eigenspace is degenerate), we write 
\begin{equation}
    T\bvph = \e^{ik_x}\bvph \implies \Phi_n = \e^{ik_x n}\bvph, 
\end{equation}
which is simply Bloch's theorem for periodic systems. 

We next set the hopping matrix connecting $\Psi_1$ and $\Psi_N \equiv \Psi_0$ as $\kappa J$ for some $\kappa \in \real$. Then, we may interpolate continuously between PBC and OBC by tuning $\kappa$ from one to zero. Following the approach of Ref.~\cite[Sec.~III.C.2]{vd_interface}, we write the modified recursion relation in Eq.~(\ref{eq:recur}) for $n=0,1$ as
\begin{align}
    \Psi_N &= \kappa \, \green J \Psi_{1} + \green J^\dg \Psi_{N-1}, \nonumber \\ 
    \Psi_1 &= \green J \Psi_{2} + \kappa \, \green J^\dg \Psi_{N}.
\end{align}
Multiplying to the left with $V^\dg$ and $W^\dg$ as earlier, these reduce for $\kappa \neq 0$ to 
\begin{equation}
    \Phi_1 = K_\R T \Phi_N, \qquad 
    \Phi_2 = T K_\L \Phi_1,
\end{equation}
respectively, where $K_\L = \diag{\id_r, \kappa\id_r}$ and $K_\R = \diag{\frac{1}{\kappa}\id_r,\id_r}$. Using $\Phi_N = T^{N-2} \Phi_2 $, we get
\begin{equation}
    \Phi_1 = K_\R T^N K_\L \Phi_1. 
    \label{eq:bc_cond}
\end{equation}
We finally set $\bvph = K_\L \Phi_1$ to obtain
\begin{equation}
    \bvph = K T^N \bvph, \qquad 
    K = \diag{\frac{1}{\kappa} \id_r, \kappa \id_r}. 
\end{equation}
Thus, we have a state \emph{iff} $1 \in \spec{K T^N(\ve, \vk_\perp)}$. For $\kappa = 1$, i.e., $K = \id_{2r}$, we recover Eq.~(\ref{eq:pbc1}), which can be reduced to a condition on the spectrum of $T$ as opposed to that of $T^N$, and can thus be readily generalized to $N\to\infty$. This is convenient, since $T^N$ is generally difficult to compute analytically. For arbitrary $\kappa$, we have been able to obtain the $N\to\infty$ limit only when $r=1$ using an explicit form of $T^N \!\!$, as described in Sec.~\ref{sec:r1_bc}.

\subsubsection{Open boundary condition}
For OBC, we need to take the limit $\kappa \to 0$, for which Eq.~(\ref{eq:bc_cond}) is singular. To remedy this, we multiply to the left by $K_\R^{-1}$ to get 
\begin{equation}
    \begin{pmatrix}
        \kappa \id_r & 0 \\ 
        0 & \id_r 
    \end{pmatrix} \Phi_1 = 
    T^N  \begin{pmatrix}
        \id_r & 0 \\ 
        0 & \kappa \id_r 
    \end{pmatrix} \Phi_1,
\end{equation}
which is well-behaved as $\kappa\to 0$. Setting $\kappa = 0$, we find 
\begin{equation}
    \vecenv{0}{\bal_N} = T^N \vecenv{\bbe_1}{0}.
    \label{eq:obc1}
\end{equation}
where $\bal_N$ and $\bbe_1$ are arbitrary. This is equivalent to the Dirichlet boundary condition used in Ref.~\cite[Sec.~II.D.3]{vd-vc_tm}, where one starts with an infinite chain and sets $\Psi_0 = \Psi_{N+1} = 0$. 

To solve this condition for $(\ve, \vk_\perp)$, the general strategy is to find solutions to the eigenvalue problem 
\begin{equation}
    T(\ve, \vk_\perp) \bvph_\ell = \rho_\ell \bvph_\ell,
\end{equation}
and to then expand $\Phi_{1}$ and $\Phi_{N+1}$ in terms of these eigenvectors. We first consider the case where $T$ is diagonalizable, so that $\bvph_\ell$ form a (generically non-orthogonal) basis of $\cmplx^{2r}$. The condition in Eq.~(\ref{eq:obc1}) then becomes
\begin{equation}
    \vecenv{\bbe_1}{0} = \sum_{\ell = 1}^{2r} a_\ell \bvph_\ell, \qquad   
    \vecenv{0}{\bal_N} = \sum_{\ell = 1}^{2r} a_\ell \rho_\ell^N \bvph_\ell. 
    \label{eq:obc2}
\end{equation}
This can be further reduced by projecting down to the sectors where the left hand side of these equations vanishes. Explicitly, 
\begin{equation}
    \sum_{\ell = 1}^{2r} a_\ell \, \proj_\bal \bvph_\ell 
    = \sum_{\ell = 1}^{2r} a_\ell \rho_\ell^N \, \proj_\bbe \bvph_\ell = 0,
\end{equation}
where the projectors $\proj_{\bal, \bbe} \cn \cmplx^{2r} \to \cmplx^r$ are defined as $\proj_\bal = \left( 0, \id_r \right)$ and $\proj_\bbe = \left( \id_r, 0 \right)$. This is a set of $2r$ complex homogeneous linear equations in $2r$ variables $\va = \{ a_1, a_2, \dots, a_{2r} \}$, which can be recast into a matrix equation of the form $\mathcal{R} \cdot \va = \nullv$, which, by Cramer's rule, has a nontrivial solution \emph{iff}
\begin{equation} 
\det \mathcal{R} = 0; \qquad 
\mathcal{R} = 
    \begin{pmatrix}
        R_1^N \bvph_1 & \dots & R_{2r}^N  \bvph_{2r}
    \end{pmatrix},
    \label{eq:obc_cramers}
\end{equation}
where we have defined
\begin{equation*}
    R_\ell = \vecenv{ \rho_\ell \proj_\bbe}{\proj_\bal} 
    = \begin{pmatrix}
        \rho_\ell \id_r & 0 \\ 
        0 & \id_r
      \end{pmatrix}.
\end{equation*} 
Since $\mathcal{R}$ is defined only in terms of the eigenvalues and eigenvectors of $T$, we obtain a condition for states that satisfy OBC purely in terms of $(\ve, \vk_\perp)$, which can be solved to get the set of energies for which the system with OBC has an eigenstate. 

On the other hand, if $T$ is non-diagonalizable or \emph{defective}, we need to augment the set of eigenvectors with the \emph{generalized eigenvectors} to form a basis of $\cmplx^{2r}$, which can then be used to expand $\Phi_0$. However, the action of the transfer matrix on these eigenvalues is more complicated than in the previous case, so that the associated conditions take the general form 
\begin{equation*}
    \vecenv{\bbe_1}{0} = \sum_{\ell = 1}^{2r} a_\ell \bvph_\ell, \quad   
    \vecenv{0}{\bal_N} = \sum_{\ell, \ell' = 1}^{2r} a_\ell f_{\ell\ell'} \bvph_{\ell'}, 
\end{equation*}
where $f_{\ell\ell'}(N)$ are products of polynomials and exponentials in $N$. In the case of $T$ diagonalizable, these reduce to $f_{\ell\ell'} = \rho_\ell^N \delta_{\ell\ell'}$.

In the following, we elucidate this idea for a simple case. Recall that if $\rho \in \spec{T}$ is a doubly degenerate eigenvalue with a single eigenvector $\bvph_1$, then the corresponding generalized eigenvector $\bvph_2$ is defined by the relations \cite{strang_book}
\begin{equation*}
    (T - \rho \id)\bvph_1 = 0, \qquad 
    (T - \rho\id)\bvph_2 = \bvph_1. 
\end{equation*}
Given $\Phi_1 = a_1 \bvph_1 + a_2 \bvph_2$ for some $a_{1,2} \in \cmplx$, the transfer matrix acts as
\begin{equation*}
    T^N \Phi_1 = \left(a_1 \rho + a_2  N\right) \rho^{N-1} \bvph_1 + a_2  \rho^N \bvph_2.
\end{equation*} 
Thus, we identify
\begin{equation*}
    \mathbf{f} = 
    \begin{pmatrix}
        \rho^N & 0 \\ 
        N \rho^{N-1} & \rho^N
    \end{pmatrix} =  \begin{pmatrix}
        \rho & 0 \\ 
        1 & \rho
    \end{pmatrix}^N,
\end{equation*}
so that $\mathbf{f}$ is the $N^{\text{th}}$ power of the Jordan normal form of $T$ in the eigenspace of $\rho$. Note that $f_{21}$ has picked up an additional term linear in $N$. In general, we may get terms that grow or decay as $N^k \rho^{N-k}$, where $k$ is the difference between the algebraic and geometric multiplicity of an eigenvalue of $T$. Thus, for OBC, the nondiagonalizability of $T$ gives rise to a family of states whose localization is not purely exponential, but has a polynomial decay. This would clearly be most apparent if the repeated eigenvalue lies on the unit circle. Another interesting case is when $\rho = 0$, where we get a state that decays to zero within a finite number of steps, independent of the system size.

\section{Restricting to rank 1}
\label{sec:r1}
In this section, we restrict the formal discussions of Sec.~\ref{sec:tm} to systems with $r=1$, which encompasses many tight-binding models of interest and has the advantage that the relevant computations are analytically tractable. Note that this condition is not directly related to either the range of hoppings or the number of degrees of freedom in a supercell; instead $r=1$ signifies that there exists a local unitary transform under which the Hamiltonian reduces to a form where the consecutive supercells are connected by a single bond.

For $r=1$, the transfer matrix $T \in \Mat(2,\cmplx)$ can be written as 
\begin{equation}
  T = \frac{1}{\xi \green_{vw}}
  \begin{pmatrix} 
    1 & \quad  - \xi \green_{ww}  \\ 
     \xi \green_{vv} & \quad \xi^2 \left( \green_{vw}\green_{wv} - \green_{vv}\green_{ww} \right) 
  \end{pmatrix},
 \label{eq:rank1_tm} 
\end{equation}
where $\green_{ab} \in \cmplx$ and $\xi \in \real^+$ is the (only) singular value of $J$. The eigenvalues of $T$ are 
\begin{equation}	
  \rho_{\pm} = \frac{1}{2} \left[ \Delta \pm \sqrt{\Delta^2 - 4 \Gamma} \right],
  \label{eq:r1_eigs}
\end{equation}
where 
\begin{align}
  \Delta  &\equiv \tr \, T = \frac{1}{\xi \green_{vw}} \left[ 1 + \xi^2 \left( \green_{vw} \green_{wv} - \green_{vv}\green_{ww} \right) \right], \nonumber \\  
  \Gamma  &\equiv \det \, T = \frac{\green_{wv}}{\green_{vw}}.
\end{align}
In Appendix~\ref{app:schur}, we show that $\green_{ab}$'s are rational functions of $\ve$, with the numerator a polynomial in $\ve$ of order $\nuc$ for $\green_{vv},\green_{ww}$ and order $\nuc-1$ for $\green_{vw},\green_{wv}$. 
We next specialize the results of Sec.~\ref{sec:tm_spec} to the present case and use them to explain various interesting aspects of non-Hermitian systems such as the skin effect and real-space EPs. We also construct a Riemann surface associated with $\ve$, which can be used to define topological invariants for the boundary states.

\subsection{Boundary conditions and spectra}
\label{sec:r1_bc}

We split this discussion between bulk and boundary spectra. For a given transverse momentum, the bulk spectrum is generically a set of closed curves in the complex plane, which can generically be written as $\ve = F(\phi)$ with $\phi \in [0,2\pi]$ and $F$ periodic in $\phi$. The boundary spectrum, on the other hand, is a discrete set of points on the complex plane.

\subsubsection{Bulk spectra}
For a system of $N$ supercells and PBC, we use Eq.~(\ref{eq:pbc1}) to write the condition for the existence of a Bloch state as 
\begin{equation} 
  \Delta = \e^{i\phi} + \Gamma \e^{-i\phi},   \label{eq:r1_pbc1}
\end{equation} 
where $\phi = 2\pi\ell/N, \, \ell \in \{0, \dots N-1\}$, and the $N\to\infty$ limit is taken by setting $\phi\in[0,2\pi)$. Since the numerator of $\Delta$ is a polynomial in $\ve$ of order $\nuc$, we obtain $\nuc$ complex solutions for $\ve$ for each $\phi$ and $\vk_\perp$. Scanning over $\phi$, we get the PBC bulk spectrum. We reiterate that if $\bvph$ is the eigenvector of $T$ associated with $\e^{i\phi}$, then the corresponding bulk states are given by $\Phi_n = \e^{i n\phi} \bvph$, which are precisely the Bloch states. 

We next turn to the condition for OBC [cf. Eq.~(\ref{eq:obc1})], which can be rewritten in the present case as 
\begin{equation}
     T^N \vecenv{1}{0} = \mathfrak{r} \vecenv{0}{1}  
     \label{eq:r1_obc1}
\end{equation}
for some $\mathfrak{r} \in \cmplx$. This can be used to derive a Cramer's condition, as was done in Sec.~\ref{sec:tm_spec}. However, for $r=1$, we can explicitly compute $T^N$ \cite{soto_tmat_geom} and use it to derive conditions involving only the transfer matrix in the $N\to\infty$ limit. As shown in Appendix~\ref{app:obc}, for $\Gamma \neq 0$, 
\begin{equation}
    T^n = \Gamma^{n/2} \left[ \frac{1}{\sqrt{\Gamma}} \, U_{n-1}(z) T - U_{n-2}(z) \id \right], 
\end{equation}
where $z = \Delta/2\sqrt{\Gamma}$ and $U_n(z)$'s are the Chebyshev polynomials of the second kind, explicitly defined in Eq.~(\ref{eq:cheby_def}). Combining this with Eq.~(\ref{eq:r1_obc1}), we derive the condition for OBC as 
\begin{equation}
    \xi \sqrt{\green_{vw} \green_{wv}} = \frac{U_{N-1}(z)}{U_{N-2}(z)}.
     \label{eq:r1_obc2}
\end{equation}
The behavior of the right hand side as $N\to\infty$ strongly depends on $z$. If $z$ is real and $z \in [-1,1]$, we set $z = \cos\phi$ for some $\phi \in [0,\pi]$ and use Eq.~(\ref{eq:cheby_def}) to rewrite Eq.~(\ref{eq:r1_obc2}) as 
\begin{equation}
    \xi \sqrt{\green_{vw} \green_{wv}} = \frac{\sin \left( N\phi\right) }{\sin \left[(N-1) \phi\right]}.
     \label{eq:r1_obc3}
\end{equation}
The right hand side has poles at $\phi = \ell \pi/(N-1)$ and zeros at $\phi = \ell\pi/N$ with $\ell = 0, 1, \dots N-1$. Thus, Eq.~(\ref{eq:r1_obc3}), solved for $\phi$, has $N$ solutions, one lying in each open set $(\ell\pi/N, \ell\pi/(N-1))$. As $N\to\infty$, we get a dense set of solutions, which is our bulk band for OBC. Thus, the condition for the bulk band in terms of $(\ve, \vk_\perp)$ can be written as 
\begin{equation}
    \Delta = 2\sqrt{\Gamma} \cos\phi
     \label{eq:r1_obc_bulk}
\end{equation}
for some $\phi\in[0,\pi]$. Since $\Delta$ is a polynomial in $\ve$ of order $\nuc$, this equation has $\nuc$ complex solutions for $\ve$ for each $\phi$. Scanning over $\phi$, we thus get the OBC bulk bands generically as a set of $\nuc$ closed curves on $\cmplx$. Furthermore, the corresponding eigenvectors can be computed analytically as 
\begin{equation}
    \Phi_n = \frac{\Gamma^{n/2}}{\sin (N \phi)} \vecenv{\sin \left( (N-n) \phi \right)}{\sqrt{\frac{\green_{vv}}{\green_{ww}}} \sin (n\phi)},
\end{equation}
which yield the correct boundary vectors for $n=0,N$. 

To obtain further insight into the meaning of this condition, we substitute Eq.~(\ref{eq:r1_obc_bulk}) in Eq.~(\ref{eq:r1_eigs}) to conclude that the eigenvalues of the transfer matrix are $\rho_\pm = \sqrt{\Gamma} \e^{\pm i\phi}$. The equality of magnitude of the eigenvalues can be alternatively inferred as follows: if $\abs{\rho_+} \neq \abs{\rho_-}$, then for any initial vector $\Phi_0$ which is a linear combination of the corresponding eigenvectors, the eigenvector corresponding to the larger magnitude of the eigenvalue will dominate as $N\to\infty$. Thus, a matching condition like Eq.~(\ref{eq:r1_obc1}) can be satisfied only if the two eigenvalues are equal in magnitude, thereby leading to Eq.~(\ref{eq:r1_obc_bulk}). This argument is essentially identical to that of Ref.~\cite{yao_nh_winding} with their decay exponents $\beta$ being  equal to the eigenvalues of the transfer matrix.

For transfer matrices with $\Gamma = 0$, excluded in the above derivation, we find $T^n = \Delta^{n-1} T$. Substituting this in Eq.~(\ref{eq:r1_obc1}) results in $\Delta = 0$, which is equivalent to $\Gamma\to 0$ limit of Eq.~(\ref{eq:r1_obc_bulk}). Thus, we get a bulk state for OBC \emph{iff}
\begin{equation}
    \Delta = \Gamma = 0 \implies \green_{wv} = 0, \quad \green_{vv} \green_{ww} = \xi \green_{vw}.
\end{equation}
Since this is independent of $\phi$ unlike Eq.~(\ref{eq:r1_obc3}), we get a discrete set of $\nuc$ points instead of $\nuc$ bands, i.e., each bulk band collapses to a single energy eigenvalue. Furthermore, 
\begin{equation}
    \Phi_1 = \vecenv{1}{0}, \quad 
    \Phi_2 = \frac{1}{\xi\sqrt{\green_{vw} \green_{wv}}} \vecenv{1}{\xi\green_{vv}},
\end{equation}
and $\Phi_n = \nullv \; \forall n > 2$. Thus, we have a single state for each band, which is localized at the left boundary and has a finite support.

The condition for the bulk states for PBC and OBC can be written concisely as
\begin{equation}
    \Delta = 2 \sqrt{\Gamma} \cos (\phi + i \zeta),
\end{equation}
with the cosine of a complex angle with $\phi \in [0,2\pi)$ and $\zeta \in \real$. For $\kappa = 0$ (OBC) and $\kappa = 1$ (PBC), we get $\zeta = 0$ and $\frac{1}{2} \log \Gamma$, respectively. We can extend these further by continuously tuning between these two values of $\zeta$ as discussed in Sec.~\ref{sec:tm_spec}. In this setup, for $0 < \kappa < 1$, we find some intermediate $\zeta = \zeta_\kappa$ that interpolates between $0$ and $\frac{1}{2} \log \abs{\Gamma}$. We derive an approximate expression for $\zeta_\kappa$ in Appendix~\ref{app:obc}. 

\subsubsection{Boundary spectra}
The boundary states are obtained as additional discrete solutions to Eq.~(\ref{eq:r1_obc2}). For $z\notin [-1,1]$, the $N\to\infty$ limit of the right hand side of Eq.~(\ref{eq:r1_obc2}) is finite, so that Eq.~(\ref{eq:r1_obc2}) reduces to the condition
\begin{equation}
    \green_{vv} \green_{ww} = 0.
     \label{eq:r1_obc_edge}
\end{equation}
The solutions to these equations give us the boundary spectrum, but more care is needed to physically interpret them. The problem stems from the fact that for $N\to\infty$, we have essentially ignored the boundary condition at the other end, thereby effectively treating the system as semi-infinite. We need to additionally ensure that the state so obtained decays into the bulk. Thus, only those solutions of Eq.~(\ref{eq:r1_obc_edge}) describe a physical left boundary mode for which the corresponding eigenvalue of the transfer matrix satisfies $\abs{\rho_\L} < 1$, and a similar condition for the right boundary mode. 

The boundary states can alternatively be obtained in a more straightforward manner by starting with a semi-infinite system and demanding that the boundary vector is an eigenvector of the transfer matrix, as in Ref.~\cite[Sec.~III.A]{vd-vc_tm}. More explicitly, a left boundary state is obtained when
\begin{equation*}
    T \vecenv{1}{0} = \rho_\L \vecenv{1}{0} \implies 
    \begin{cases}
         \green_{vv} = 0, & \\ 
         \rho_\L = \left(\xi \green_{vw}\right)^{-1}. & 
    \end{cases}
\end{equation*}
A similar calculation for the right boundary results in $\green_{ww} = 0$ and $\rho_\R = \xi \green_{wv}$, in agreement with Eq.~(\ref{eq:r1_obc_edge}). We can alternatively write the expressions for boundary spectra as a special case of the equation
\begin{equation} 
    \bvph^T \, \symp \, T(\ve, \vk_\perp) \, \bvph_0 = 0; \qquad 
    \symp = \begin{pmatrix}
        0 & 1 \\ 
        -1 & 0
        \end{pmatrix},   
    \label{eq:evans_cond}
\end{equation} 
since $\bvph^T \symp \bvph = 0, \, \forall  \bvph \in \cmplx^2$. 
Setting $\bvph = (1,0)^T$ or $(0,1)^T$, we recover the boundary state conditions computed above. In writing this equation, we have ignored the decay condition, so that we obtain physical states (in ED, for instance) only for a subset of the solutions of Eq.~(\ref{eq:evans_cond}). On the other hand, a solution to this equation exists for all $\vk_\perp$. For two-dimensional systems where $\vk_\perp \in S^1$, this fact can be used to define closed curves corresponding to the boundary states on a Riemann surface, as we show in Sec~\ref{sec:r1_rsurf}.

\subsection{Aspects of non-Hermiticity}
\label{sec:r1_prop}

We now discuss several intriguing aspects of non-Hermitian systems that can be readily deduced from the knowledge of its transfer matrix. 

\begin{figure}
  \includegraphics[width=0.9\columnwidth]{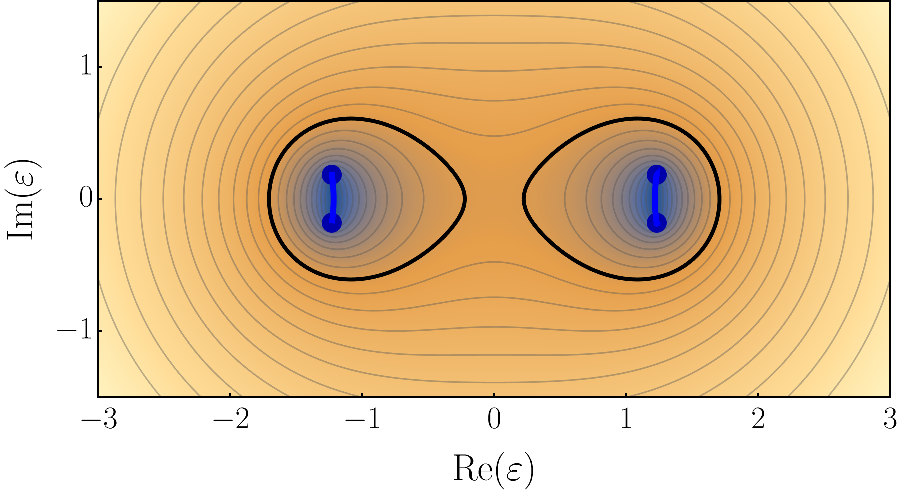} \\ \includegraphics[width=0.9\columnwidth]{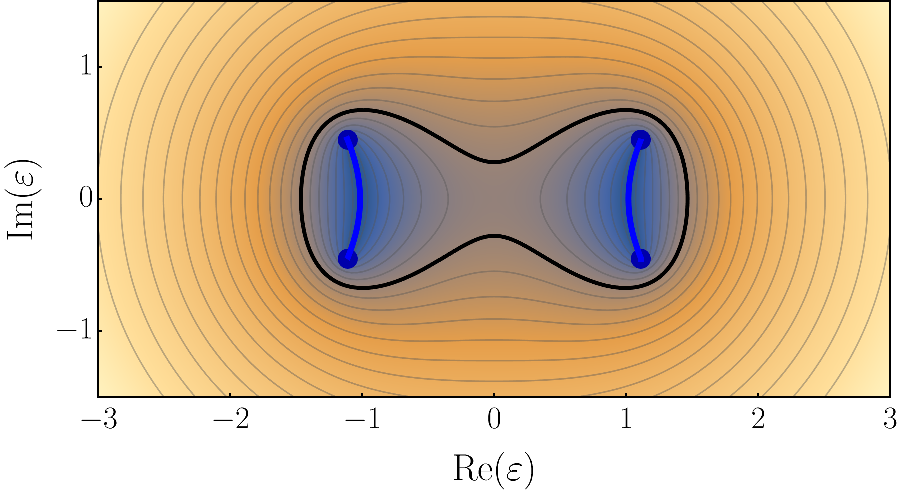} 
  \caption{$\log \abs{\rho(\ve)}$ as a function of complex $\ve$, with positive (negative) values indicated by yellow (blue). The dark blue lines are the locii where $\abs{\rho_+} = \abs{\rho_-} = \sqrt{\abs{\Gamma}}$, along which we get the OBC bulk band, while the black lines correspond to $\abs{\rho} = 1$, along which we get the PBC bulk band. These plots are computed for the model in Sec.~\ref{sec:eg_ci_hy} with the parameters corresponding to those in the right column of Fig.~\ref{fig:ci_hy_imag} with $k_y = 0.26\pi$ (top) and $0.18\pi$ (bottom).}
  \label{fig:evals}
\end{figure}

\subsubsection{PBC vs OBC bulk spectra}
\label{sec:r1_gamma1}
The difference between the PBC and OBC bulk spectra can be easily visualized by plotting the magnitudes of the eigenvalues of the transfer matrix, as shown in Fig.~\ref{fig:evals}. The bulk bands for PBC are then given by the intersection of the plane $\abs{\rho} = 1$ with the eigenvalues (black lines in Fig.~\ref{fig:evals}), while those for OBC are given by the intersection with the plane where the two eigenvalues are equal in magnitude, i.e., $\abs{\rho} = \sqrt{\abs{\Gamma}}$ (blue lines). For $\abs{\Gamma} \neq 1$, these two planes do not coincide, so that their intercepts, i.e., the curves on the complex $\ve$ plane corresponding to the bulk bands, can be different in the two cases. 

On the other hand, these two curves become identical if $\abs{\Gamma} = 1$. We can also see this analytically, since setting $\Gamma = \e^{-2i\chi}$ for some $\chi \in [0,2\pi)$, the conditions for PBC and OBC bulk modes in Eqs.~(\ref{eq:r1_pbc1}) and (\ref{eq:r1_obc_bulk}) both reduce to
\footnote{ 
    Here we have replaced $(\phi-\chi)$ with $\phi$ in the condition for PBC [cf. Eq.~(\ref{eq:r1_pbc1})] since we are scanning over $\phi$.
}
\begin{equation}
    \Delta = 2 \, \e^{-i\chi} \, \cos (\phi),
    \label{eq:unimod_obc}
\end{equation}
with $\phi \in [0,2\pi)$. Thus, in the large system limit, the bulk spectra for a system with PBC and OBC are identical \emph{iff} the transfer matrix is unimodular. In Sec.~\ref{sec:tm_spcl}, we showed that Hermiticity or $\PT$-symmetry implies unimodularity of the transfer matrices for physically relevant energies. This may, however, also be true in more general settings. 

This difference between the PBC and OBC bulk spectra can lead to an interesting situation, where as one tunes a parameter (or $k_y$), the PBC bulk gap closes while the OBC bulk bands remain gapped. This scenario is depicted in Fig.~\ref{fig:evals}. In this case, any topologically nontrivial boundary states, if present, will also remain qualitatively unchanged, since they cannot be removed without closing the gap between the two bands connected by them, i.e., the OBC bulk bands. Thus, this presents an instance of a dramatic breakdown of the conventional bulk-boundary correspondence. 

\subsubsection{The non-Hermitian skin effect}
To study the skin effect, we need to look at the asymptotic behavior of the states for systems with PBC and OBC. For systems with PBC, 
\begin{equation*}
    \norm{\Phi_n} = \norm{\e^{i n \phi} \bvph_1} = \norm{\bvph_1}
\end{equation*}
independent of $n$, as one would expect for Bloch waves. On the other hand, for OBC, we have 
\begin{equation*}
    \norm{\Phi_n} = \norm{\Gamma^{n/2} \left( a_1 \e^{i n \phi} \bvph_1  + a_2 \e^{-i n \phi} \bvph_2 \right) } \sim \abs{\Gamma}^{n/2}. 
\end{equation*}
If $\abs{\Gamma} \neq 1$, the ``bulk states'', or more precisely, the states associated with the continuum spectrum for the system with PBC, decay into the bulk. These states are localized on the left boundary for $\abs{\Gamma} < 1$ and on the right boundary for $\abs{\Gamma} > 1$. Thus, the existence of the \emph{non-Hermitian skin effect} can be deduced simply from the value of $\abs{\det T}$. 

Combining this with the result from the previous subsection, we note that the phenomena of the skin effect and the difference between the PBC and OBC bulk spectra are intimately linked, since they are both governed by the same condition. More explicitly, a non-Hermitian system exhibits the skin effect \emph{iff} the PBC and OBC bulk spectra are different.

\subsubsection{Exceptional points}
The Bloch and real-space EPs can also be understood in the transfer matrix formalism. We have a Bloch EP if the condition for the bulk states, i.e., Eq.~(\ref{eq:r1_pbc1}), solved for $\ve$, has a repeated root. The multiplicity of the roots sets the order of the EP. On the other hand, the real-space EPs are obtained when $\abs{\Gamma} \to 0, \infty$. We remark that the order of the real-space EP is $(N-1)$, where $N$ is the system size, so that we can get EPs with arbitrarily high order for a given Bloch Hamiltonian. On the other hand, the order of the Bloch EPs is limited by the dimensionality of the Bloch Hamiltonian. Thus, if $\abs{\Gamma} \neq 0, \infty$, then the maximum order of an EP in the real-space spectrum is restricted by the dimensionality of the Bloch Hamiltonian, where we make use of the fact that when $\Gamma \neq 0, \infty$ the a nonunitary similarity transform of the original Hamiltonian yields a Hamiltonian for which $\Gamma = 1$ \cite{yao_nh_winding}.

\subsubsection{Bulk topological invariants}
A bulk-boundary correspondence for non-Hermitian gapped systems can be defined if one computes the ``bulk invariant'' using the eigenvectors for a system with boundaries as opposed to that for a periodic system. In the present setup, the eigenvectors for PBC vary with position as $\e^{i n \phi}$ where $\phi$ can be identified with $k_x$ while those for OBC go as $\sqrt{\Gamma} \e^{i n \phi}$. This suggests the topological invariants should be computed using a modified Bloch Hamiltonian with the replacement 
\begin{equation}
    \e^{i k_x} \to \sqrt{\Gamma} \e^{i\phi} \iff 
    k_x \to \phi - \frac{i}{2} \log\Gamma.
\end{equation}
For instance, for gapped two-dimensional systems, a modified Chern number can be computed by integrating the biorthogonal Berry curvature on the modified ``Brillouin zone'' with coordinates $(\phi, k_y)$. 

This approach was indeed shown to predict the existence of edge modes correctly in Refs.~\cite{yao_nh_winding, yao_nh_ci}, albeit only for specific tight-binding models. Our setup thus provides a direct way of analytically generalizing the topological invariants for Hermitian Hamiltonians to non-Hermitian Hamiltonians to arbitrary lattice models without resorting to continuum limits or numerical computations.

\subsubsection{Biorthogonality condition} 
The case of rank-1 systems subsumes the non-Hermitian models discussed in Ref.~\cite{flore_biorth}, whose boundary modes can be obtained analytically by construction. As an indicator of the existence of boundary modes, a \emph{biorthogonal polarization} was proposed, defined in terms of $\mathfrak{p} \equiv \abs{\wt{\rho}_\L^\ast \rho_\L^{\phantom{\ast}}}$, i.e., the product of decay exponents of the left and right eigenstates of the Hamiltonian, localized at the left boundary. It was shown that the boundary states merge into the bulk band when $\abs{\mathfrak{p}} = 1$. 

We now derive this quantity using the transfer matrix formalism by considering a semi-infinite non-Hermitian system on $\intg^+$. Let $\Psi$ be the right eigenstate of the Hamiltonian for a left boundary mode, with the decay exponent $\rho_\L = -[\xi \green_{vw}(\ve_\L)]^{-1}$, where $\ve_\L$ satisfies $\green_{vv}(\ve_\L) = 0$. For the corresponding left eigenstate, we need the transfer matrix $\wt{T}$ for $\wt{\hlt} = \hlt^\dg$ in terms of which the decay exponent is given by $\wt{\rho}_\L = -[\xi \wt{\green}_{vw}(\ve_\L)]^{-1}$. Using Eq.~(\ref{eq:tm_left_eig}) to relate $\green$ to $\wt{\green}$, we find 
\begin{equation}
    \wt{\rho}_\L = -\frac{1}{\xi \green^\ast_{wv}(\ve_\L^\ast)} 
    \implies \mathfrak{p}  = \frac{1}{\xi^2  \abs{ \green_{vw} \green_{wv}} }.
\end{equation}
Next, we note that the bulk and boundary bands merge for a given $(\ve, \vk_\perp)$ if the conditions for both bulk and boundary states for OBC are simultaneously satisfied. Thus, we seek to solve $\green_{vv} = 0 = \Delta - 2\sqrt{\Gamma} \, \cos \phi$ for some $\phi$. We combine these to get
\begin{equation}
    \frac{1 + \xi^2 \green_{vw} \green_{wv}}{\xi \green_{vw}} = 2\sqrt{\frac{\green_{wv}}{\green_{vw}}} \cos\phi,
\end{equation}
which can be rearranged as
\begin{equation*}
    \mathfrak{p} - 2 \sqrt{\mathfrak{p}} \, \cos\phi  + 1 = 0.
\end{equation*}
This is solved by $\sqrt{\mathfrak{p}} = \e^{\pm i\phi}$, which is equivalent to demanding that $\abs{\mathfrak{p}} = 1$, precisely what was obtained in Ref.~\cite{flore_biorth}. Note that the exact same condition is obtained by alternatively considering $\abs{\wt{\rho}_\R^\ast \rho_\R^{\phantom{\ast}}}$ for the right boundary.

\subsection{\texorpdfstring{$\ve$-}{}Riemann surface}
\label{sec:r1_rsurf}
The algebraic structure of the transfer matrix naturally lends itself to the construction of a Riemann surface associated with the complex energy. Explicitly, the map $\ve \mapsto \rho$ of in Eq.~(\ref{eq:r1_eigs}) is not analytic for $\ve\in\cmplx$, since there are square-root singularities at the zeros of $Q(\ve) \equiv \Delta^2(\ve) - 4\Gamma(\ve)$. Since $\Delta$ and $\Gamma$ are rational functions in $\ve$, so is $Q(\ve)$, with the numerator being a polynomial of order $2\nuc$ (see Appendix \ref{app:schur} for details). Thus, $Q(\ve)$ has exactly $2\nuc$ complex roots, which must be connected in pairs by $\nuc$ branch cuts. Since these zeros are points where the two eigenvalues of the transfer matrix are degenerate, i.e., where 
\[ 
\rho_+ = \rho_- = \Delta/2 = \pm \sqrt{\Gamma}, 
\]  
we define the branch cuts to lie along the bulk spectrum for OBC. More explicitly, we define the branch cuts as the curves in the $\ve$ plane for which $\rho_\pm = \sqrt{\Gamma} \e^{\pm i\phi}$. For example, in Fig.~\ref{fig:evals}, we have $\nuc = 2$, and the four zeros of $Q(\ve)$ are denoted by dark blue dots, with the branch cuts lying along the blue solid lines.

Two copies of $\cmplx$ are glued along these branch cuts and a compact Riemann surface $\rsurf$ is then obtained by one-point compactifying these sheets into Riemann spheres and gluing them. By the Riemann-Hurwitz lemma, we deduce that $\rsurf$ has genus $(\nuc - 1)$, i.e., one less than the number of Bloch bands. Thus, in the case of Fig.~\ref{fig:evals}, the Riemann surface is a 2-torus. An explicit mapping from the $\ve$-plane with two branch cuts to a torus can be implemented via the elliptic integrals, as shown in Ref.~\cite[Sec.~IV.A]{vd-vc_tm}. 

This construction particularly caters to a system with OBC. For each $\vk_\perp$, the continuum states run precisely along the branch cuts. Furthermore, for a two-dimensional system where $\vk_\perp = k_y \in S^1$, the boundary modes are a map $S^1 \to \rsurf$, which can be classified by a winding number. This is the topological invariant associated with the boundary states. This approach can be used to define a ``topologically protected boundary mode'' for non-Hermitian systems as modes with nonzero winding numbers, in absence of the conventional definition for Hermitian systems in terms of directed zero crossings.

\subsection{A generic two-band model}
\label{sec:r1_model}

We now illustrate the ideas discussed in this section by explicit computations on a generic two-band model. We consider a $d$-dimensional system described by an arbitrary Bloch Hamiltonian of the form
\begin{equation} 
  \hlt_{\text{B}}(k_x, \vk_\perp) = \hlt_0(k_x) + \bet(\vk_\perp)  \cdot \bsg, \label{eq:bloch_ham_two_band_model}
\end{equation} 
where $\bet \cn \torus^{d-1} \to \cmplx^3$ depends on $\vk_\perp$, $\bsg = \left( \sigma^x, \sigma^y, \sigma^z \right)$ is the vector of Pauli matrices, and
\begin{equation} 
  \hlt_0(k_x) = \cos k_x \, \sigma^x - \sin k_x \, \sigma^y = 
  \begin{pmatrix} 
    0 & \e^{i k_x} \\ 
    \e^{-i k_x} & 0 
  \end{pmatrix}.
\end{equation} 
The eigenvalues of the Bloch Hamiltonian are
\begin{equation}
    \ve = \pm \left[ 1  + \eta^2 + 2 (\eta_x \cos k_x - \eta_y \sin k_x) \right]^{1/2}, \label{eq:eigs_bloch}
\end{equation}
where 
\begin{equation*}
  \eta^2 \equiv \bet \cdot \bet = \bet_R \cdot \bet_R -  \bet_I \cdot \bet_I + 2 i \, \bet_R \cdot \bet_I,
\end{equation*}
with $\bet_{R}$ and $\bet_{I}$ the real and imaginary parts of $\bet$, respectively. Note that $\eta$ here is not the usual norm of $\bet \in \cmplx^3$, i.e., $\eta^2 \neq \bet \cdot \bet^\ast$.

To compute the transfer matrix, we identify the hopping and on-site matrices as coefficients of $\e^{i k_x}$ and $1$ in the Bloch Hamiltonian, so that
\begin{equation} 
  J = \begin{pmatrix}
       0 & 1 \\ 
       0 & 0
      \end{pmatrix}, \qquad 
  M = \bet \cdot \bsg.
\end{equation}
The SVD results in $J = \xi \, \vv \cdot \vw^\dg$, with
\begin{equation} 	
  \vv = \vecenv{1}{0}, \qquad 
  \vw = \vecenv{0}{1}, \qquad 
  \xi = 1.
\end{equation}
The on-site Green's function is
\begin{equation} 
  \green = \left( \ve \id - \bet \cdot \bsg \right)^{-1} = \frac{\ve \id + \bet \cdot \bsg}{\ve^2 - \eta^2}.
\end{equation} 
Writing $\green$ as a matrix for the given definitions of $\vv$ and $\vw$, we identify 
\begin{equation} 
    \begin{pmatrix} 
      \green_{vv} & \green_{wv} \\ 
      \green_{vw} & \green_{ww} 
    \end{pmatrix}
  = \frac{1}{\ve^2 - \eta^2} 
  \begin{pmatrix}
    \ve + \eta_z & \eta_x - i \eta_y \\ 
    \eta_x + i \eta_y & \ve - \eta_z 
  \end{pmatrix}. 
\end{equation} 
Using Eq.~(\ref{eq:rank1_tm}), the transfer matrix is
\begin{equation} 
  T(\ve, \vk_\perp) = \frac{1}{ \eta_x + i \eta_y } 
  \begin{pmatrix}
    \ve^2 - \eta^2 & - \ve - \eta_z  \\ 
    \ve - \eta_z & -1
  \end{pmatrix}. \label{eq:rank1_tm_two_band_mod}
\end{equation}
We compute 
\begin{align} 
  \Delta = \frac{\ve^2 - \eta^2 - 1}{\eta_x + i \eta_y}, \qquad 
  \Gamma = \frac{\eta_x - i \eta_y}{\eta_x + i \eta_y}.
\end{align} 
in terms of which the eigenvalues of $T$ are given by \eq{eq:r1_eigs}. 

For PBC, the energies of Bloch states are given by Eq.~(\ref{eq:r1_pbc1}), which can be simplified to get 
\begin{align}
    \ve^2 = 1 + \eta^2 + 2 \, [\eta_x \cos\phi - \eta_y \sin\phi].
    \label{eq:r1model_pbc}
\end{align}
We note that this expression is identical to Eq.~(\ref{eq:eigs_bloch}) with the identification $\phi \to k_x$, as expected. For OBC, the bulk states are given by Eq.~(\ref{eq:r1_obc1}), which simplifies to
\begin{align}
    \ve^2 = 1 + \eta^2 + 2 \cos\phi \sqrt{\eta_x^2 + \eta_y^2}.
    \label{eq:r1model_obc}
\end{align}
These states are localized near the left boundary if $\abs{\Gamma} < 1$, i.e, if 
\begin{align}
    \abs{\eta_x - i\eta_y}^2 < \abs{\eta_x + i\eta_y}^2 \implies \text{Im}[\eta_x^\ast \eta_y] < 0.
    \label{eq:obc_bulk_loc}
\end{align}
Similarly, they are localized near the right boundary if $\abs{\Gamma} > 1$, i.e, if $\text{Im}[\eta_x^\ast \eta_y] > 0$.

When $\Gamma = 0, \infty$, i.e., $ \eta_x = \pm i\eta_y$, we get the real-space EP, where the bulk bands collapse to two points with energies $\ve = \pm \sqrt{1 + \eta_z^2}$. The corresponding states are all localized on the leftmost/rightmost supercell for $\eta_x = \pm i\eta_y$. Since $\eta_x \pm i\eta_y$ is the intracell hopping, these real-space EPs occur when the system has a completely unidirectional hopping, so that all states are piled up at one end of the system. Finally, we note that Eqns.~(\ref{eq:r1model_pbc}) and (\ref{eq:r1model_obc}) become identical if the transfer matrix is unimodular, as follows from Eq.~(\ref{eq:unimod_obc}). 

The boundary states are given by \eq{eq:evans_cond}, so that the boundary spectra and the corresponding decay exponents become
\begin{align} 
  \ve_\L &= -\eta_z, \qquad \rho_\L = -(\eta_x + i\eta_y)^{-1}, \nonumber \\  
  \ve_\R &= \eta_z, \qquad \;\;\, \rho_\R = -(\eta_x - i \eta_y).
  \label{eq:bdry_spec}
\end{align}
The left boundary state exists for $\vk_\perp$ where $\abs{\eta_x + i\eta_y} > 1$, while the right one exists if $\abs{\eta_x - i\eta_y} > 1$. Using this boundary spectrum, we can also compute 
\begin{equation}
    \mathfrak{p} = \abs{ \frac{\eta_z^2 - \bet^2}{\eta_x^2 + \eta_y^2} } =  \eta_x^2 + \eta_y^2, \label{eq:biorth_pol_two_dim_ham}
\end{equation}
which signals that the boundary states merge into the bulk bands for $\abs{\mathfrak{p}} = 1$, i.e., for $\abs{\eta_x^2 + \eta_y^2} = 1$. This is identical to the result obtained from the decay exponents.

To illustrate these transfer-matrix calculations, we briefly discuss a well-known example of the one-dimensional Su-Schrieffer-Heeger (SSH) model \cite{ssh}, whose various non-Hermitian variations have been studied in the literature \cite{schomerus_defect_modes, xioing_nh_bulk-bdry, yao_nh_winding, lee_half_winding, flore_biorth, shu_winding_geom, bardarson_svd_inv, lee_anatomy_of}. We consider a non-Hermitian SSH model with an asymmetry between the left and right intra-cell hoppings \cite{flore_biorth, shu_winding_geom}. Thus, explicitly we consider the Bloch Hamiltonian \begin{align}
    \hlt_{\text{Bloch}} = (\cos k_x + t) \sigma^x + (\sin k_x + i\gamma) \sigma^y.
\end{align}
As shown in Ref.~\cite{flore_biorth}, conventional bulk-boundary correspondence is broken for this model, and the bulk states pile up at the ends. We now show how the same result is derived using the transfer matrix. 

Since the Bloch Hamiltonian takes the same form as Eq.~(\ref{eq:bloch_ham_two_band_model}), we readily identify $\bet = (t, i \gamma,0)$. Using Eq.~(\ref{eq:rank1_tm_two_band_mod}), we get the transfer matrix 
\begin{equation*} 
  T(\ve) = \frac{1}{t - \gamma} 
  \begin{pmatrix}
    \ve^2 - t^2 + \gamma^2 & - \ve  \\ 
    \ve & -1
  \end{pmatrix},
\end{equation*}
so that
\begin{equation*}
\Delta = \frac{\ve^2 - t^2 + \gamma^2 - 1}{t - \gamma}, \qquad \Gamma = \frac{t+\gamma}{t-\gamma}.
\end{equation*}
Thus, the bulk spectra are given by
\begin{align*}
\ve_{\textrm{PBC}}^2 &= t^2 - \gamma^2 +1 +2 t \cos \phi - 2 i \gamma \sin \phi, \\
\ve_{\textrm{OBC}}^2 &= t^2 - \gamma^2 +1 +2 \cos \phi \sqrt{t^2 - \gamma^2}.
\end{align*}
The two bulk spectra are different, i.e., $\ve_{\textrm{PBC}} \neq \ve_{\textrm{OBC}}$ for $\gamma \neq 0$. The system exhibits the non-Hermitian skin effect, since the bulk states for OBC are localized at the left/right end when $t \gamma \lessgtr 0$, as follows from Eq.~(\ref{eq:obc_bulk_loc}). The transition between these two cases occurs at $\gamma = 0$, where the system is Hermitian and hence bulk states extend across the system. Finally, we get real space EPs for $t = \pm \gamma$, where the bulk bands collapse to $\ve = \pm 1$. 

Moreover, from Eq.~(\ref{eq:bdry_spec}), we find that the system exhibits boundary modes in the gap with energies $\ve_\L = \ve_\R = 0$ and decay exponents $\rho_\L = -1/(t - \gamma)$ and $\rho_\R = - (t+\gamma)$. The biorthogonal polarization in $\mathfrak{p}$ in Eq.~(\ref{eq:biorth_pol_two_dim_ham}) then reads as $\mathfrak{p} = t^2 - \gamma^2$ such that the boundary states attach to the bulk bands when $t = \pm \sqrt{\gamma^2 +1}, \pm \sqrt{\gamma^2 -1}$, in accordance with the results in Ref.~\cite[Eq.~(9)]{flore_biorth}. Finally, we note that the associated bulk invariant can be computed as the Berry phase computed around a modified Brillouin zone, which is obtained by the replacement
\begin{equation*}
    k_x \rightarrow \phi - \frac{i}{2} \log \Gamma = \phi - \frac{i}{2} \log \left( \frac{t+\gamma}{t-\gamma} \right).
\end{equation*}
This expression corresponds exactly to the result derived in Ref.~\cite{yao_nh_winding}. 
 
% We note that the solution for $\ve_{\textrm{PBC}}$ corresponds to the Bloch spectrum of the Bloch Hamiltonian $\hlt_{\text{B}}(k_x)$ when identifying $\phi$ with the Bloch momentum $k_x$.

\section{Two-Dimensional Examples}
\label{sec:eg}

We now apply the ideas discussed above to a variety of lattice models for two-dimensional topological phases. The PBC and OBC spectra are computed analytically and compared to those computed using numerical exact diagonalization (ED). We also discuss the topology associated with the boundary states in terms of the energy Riemann surface. 

% We discuss two instances of the generic two-band model discussed in Sec.~\ref{sec:r1_model}, as well as a non-Hermitian generalization of the Hofstadter model.  

\subsection{A non-Hermitian Chern insulator}

We consider a non-Hermitian generalization of the two-dimensional Chern insulator \cite{yao_nh_winding, yao_nh_ci, ueda_nh_ci}, for which we take
\begin{equation} 
    \bet(k_y) = \left(\cos k_y - m, 0, \sin k_y \right) + i \nhh,
    \label{eq:hlt_ci}
\end{equation} 
where $\nhh = (h_x, \, h_y, \, h_z) \in \real^3$. Physically, $h_x$ and $h_y$ represent an anisotropy in the phase and amplitude of the intracell left and right hoppings, respectively, while $h_z$ represents an on-site gain and loss on alternative sublattices. 

\begin{figure}
  \includegraphics[width=0.9 \columnwidth]{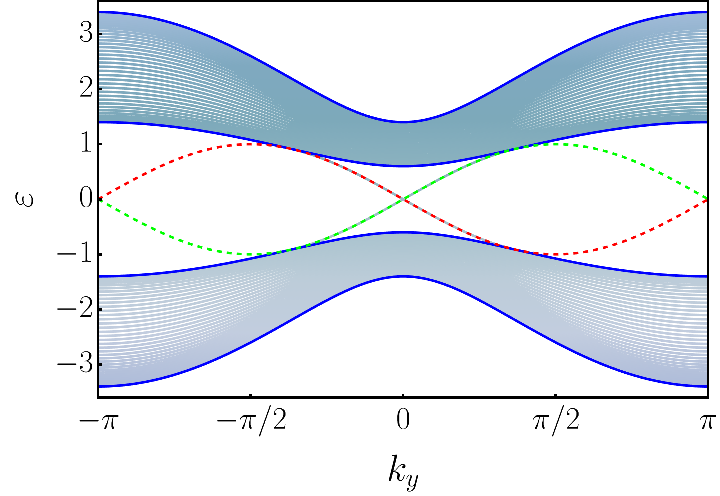}
  \caption{ 
      The spectrum of the Hermitian Chern insulator computed using numerical ED with $N=40$ and $m=1.4$ with the band edges (blue, solid), and the left (green, dashed) and right edge states (red, dashed) computed analytically using the transfer matrix. 
  }
\label{fig:ci_hermitian}
\end{figure}

For $\nhh = \nullv$, i.e., the Hermitian limit, the system is gapless for $m = 0, \pm 2$, a trivial insulator for $\abs{m}>2$, and a topological insulator with Chern number $\pm 1$ for $\abs{m}<2$. For OBC along $x$, the topological phase exhibits modes localized on the edge. Using the transfer-matrix method, we can compute the edge spectra as $\ve_{\L, \R} = \mp \sin k_y$, with the corresponding decay exponents being $\rho_\L = \cos k_y - m$ and $\rho_\R = 1/(\cos k_y - m)$, respectively \cite[Sec.~III.D.3]{vd-vc_tm}. Demanding that the edge modes decay into the bulk, we deduce that they exist near $k_y = 0$ for $0<m<2$ and near $k_y = \pi$ for $-2<m<0$. In the following analysis, we set $m=1.4$, for which we get the celebrated Chern insulator spectrum, as shown in Fig.~\ref{fig:ci_hermitian}.  

\begin{figure*}[t]
  \includegraphics[width=0.7\columnwidth]{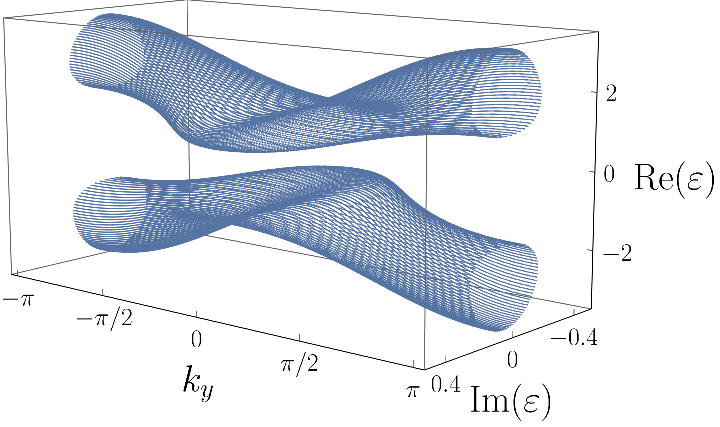} 
  \includegraphics[width=0.7\columnwidth]{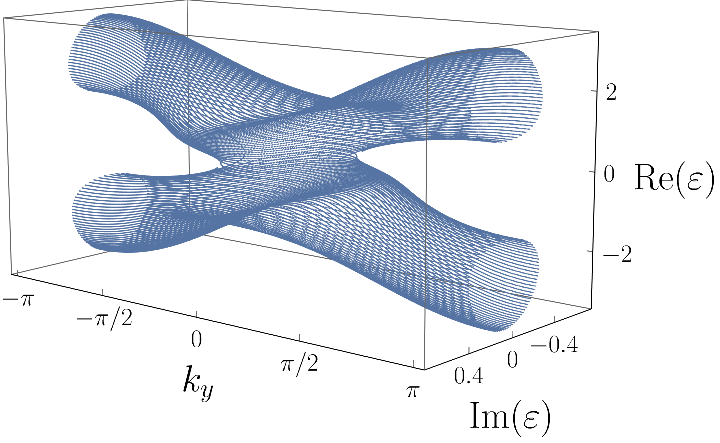} 
  \includegraphics[width=0.55\columnwidth]{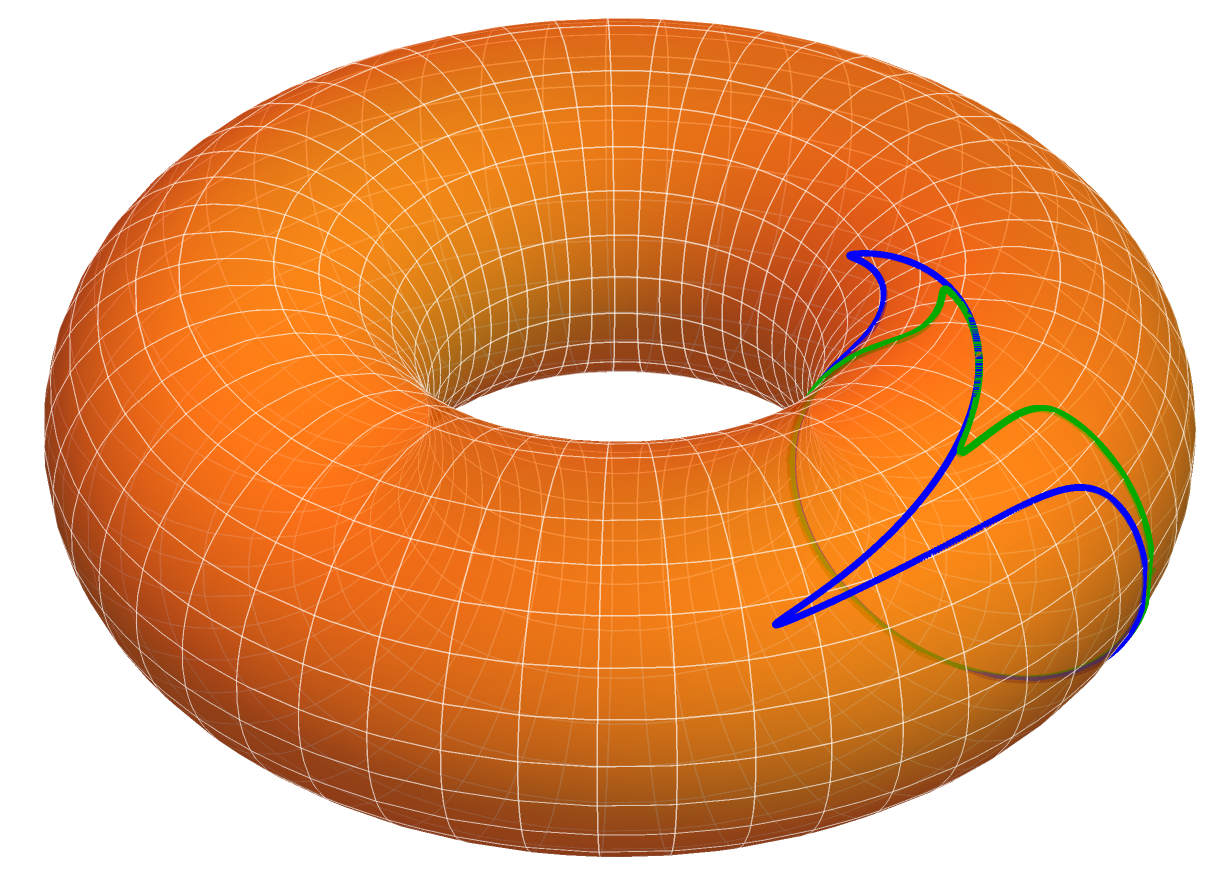} 
  \caption{Bulk bands for the non-Hermitian Chern insulator with PBC for $N=80$, $m = 1.4$, and $h_y = 0.5$ (left) and $0.75$ (middle) for PBC. The topology of the surface traced out by the bulk bands as a function of $k_y$ changes as a function of $\gamma$. The right panel shows the $\ve$-Riemann surface with the left edge states for these two cases (plotted in green and blue, respectively). Both of these wind around the same noncontractible loop and they are clearly unaffected by the PBC bulk band topology.
  }
\label{fig:ci_hy_3d}
\end{figure*}

For the non-Hermitian generalization, we find 
\begin{align}
    \Delta &= \frac{\ve^2 - ( \cos k_y - m + ih_x)^2 + h_y^2 - (\sin k_y + ih_z)^2 - 1}{  \cos k_y - m  + ih_x - h_y}, \nonumber \\ 
    \Gamma &= \frac{ \cos k_y - m + ih_x + h_y}{ \cos k_y - m  + ih_x - h_y}.
    \label{eq:tr_T_hx_imag}
\end{align}
Thus, the transfer matrix is unimodular if $h_y=0$ and non-unimodular otherwise. Since these two cases exhibit qualitatively different behaviors, we shall distinguish between them in the following analysis. with the OBC bulk states localized on the left/right edge when
\begin{equation}
    (\cos k_y - m) h_y \lessgtr 0,    
\end{equation}
as follows from Eq.~(\ref{eq:obc_bulk_loc}). Thus, the system exhibits the non-Hermitian skin effect only when the non-Hermiticity is along $\sigma^y$, corresponding to asymmetric hopping within the unit cell. Furthermore, for a fixed $h_y$ and $\abs{m}<1$, there are OBC bulk states localized at both ends of the system, corresponding to different ranges of $k_y$. 

The bulk states can be computed explicitly from Eqs.~(\ref{eq:r1_pbc1}) and (\ref{eq:r1_obc_bulk}). The edge states occur at energies $\ve_{\L, \R} = \mp (\sin k_y + ih_z)$, with the associated decay exponents $\rho_\L = \eta_x + i\eta_y$ and $\rho_\R = 1/(\eta_x - i\eta_y)$, respectively. We now set the terms in $\nhh$ to $\gamma\in\real^+$ one by one and apply the results of Sec.~\ref{sec:r1_model} to deduce the behavior of the OBC spectrum.

\subsubsection{Non-unimodular transfer matrix}
\label{sec:eg_ci_hy}
We begin with the most interesting case, \viz., that with a non-unimodular transfer matrix, by setting $h_y = \gamma$. In this case, the system exhibits the non-Hermitian skin effect as well as a difference in the PBC and OBC spectra. Interestingly, in certain parameter ranges, the PBC spectrum is actually gapless, while the OBC spectrum remains gapped with a robust edge mode in the gap (as also pointed out in Ref.~\cite{ueda_nh_ci} from ED.). The robustness of the edge mode is clear from its winding on the energy Riemann surface, which remains unchanged throughout this transition, as shown in Fig.~\ref{fig:ci_hy_3d}. We also visualize the transition in the PBC spectrum by plotting the complex PBC bulk spectrum as a function of $k_y$, which forms a surface whose topology changes from two cylinders to a ``pair-of-pants''.

\begin{figure*}[t]
  \centering
  \includegraphics[width=\columnwidth]{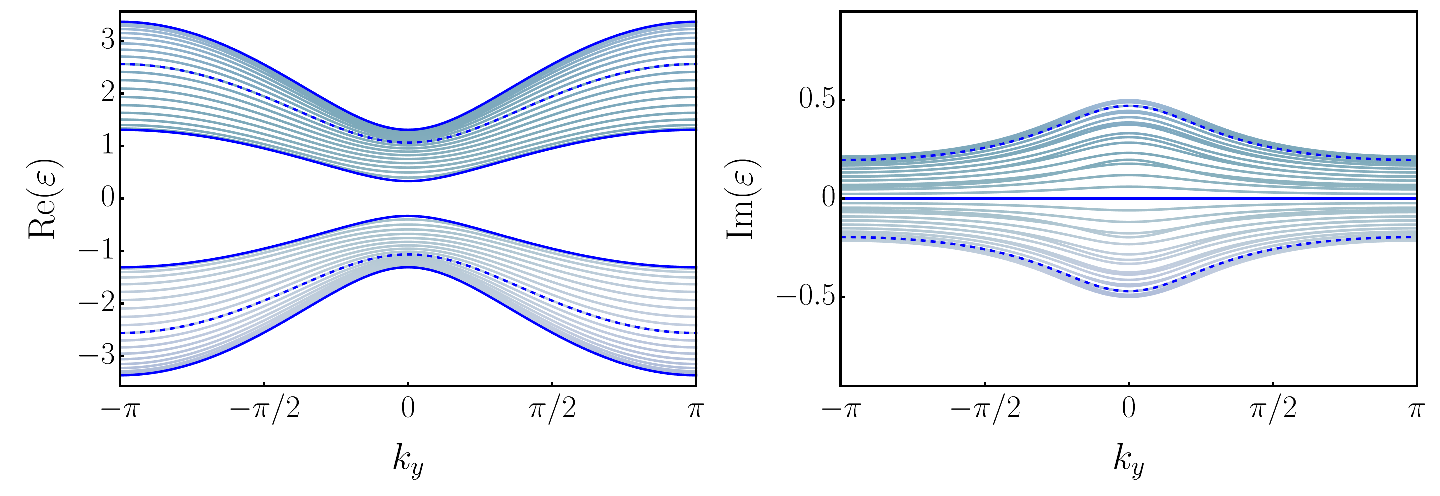}
  \includegraphics[width=\columnwidth]{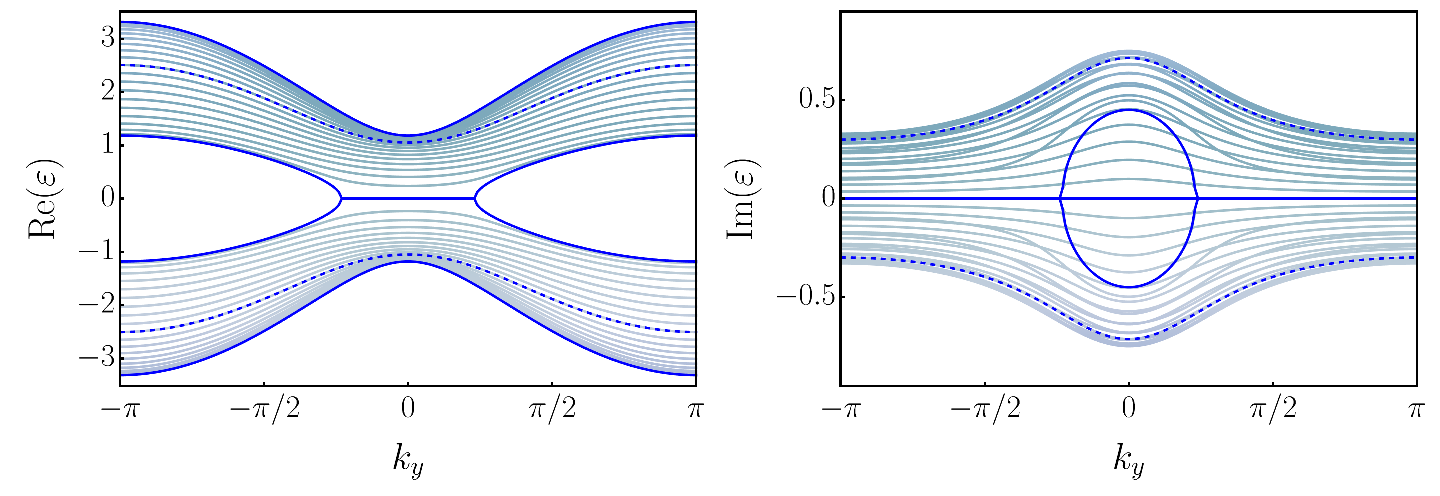} \\
  \includegraphics[width=\columnwidth]{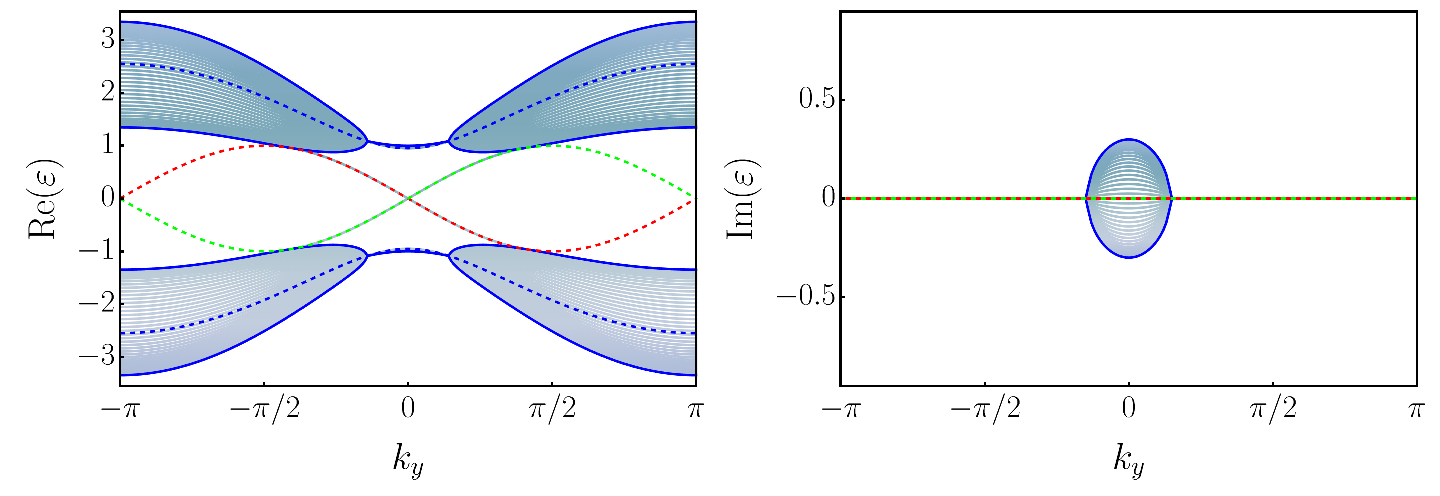} 
  \includegraphics[width=\columnwidth]{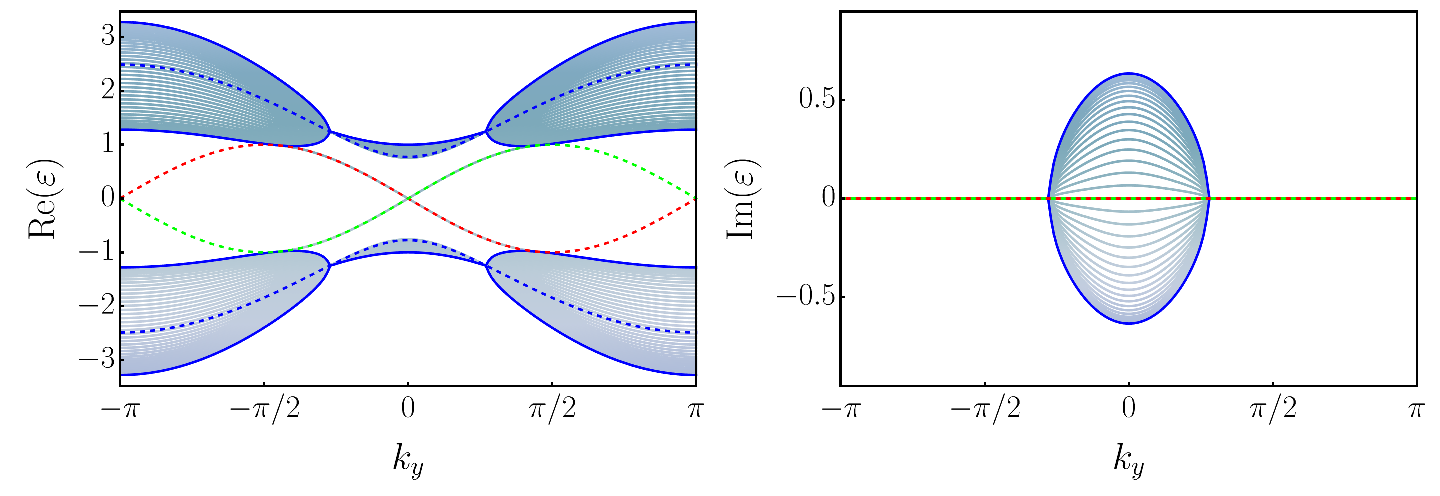}
  \caption{Analytically and numerically computed real and imaginary band structures for the Chern insulator with $N=40$, $m = 1.4$ and $h_y = 0.5$ (left column) and $0.75$ (right column) for PBC (top) and OBC (bottom). We note the qualitative difference between the PBC and OBC bulk spectra in both cases. Furthermore, the former case exhibits only real-space EPs, while the latter exhibits both real-space and Bloch EPs, but for different values of $k_y$.
  }
\label{fig:ci_hy_imag}
\end{figure*}

The bulk spectra for PBC and OBC are given by 
\begin{align*}
    \ve^2_{\text{PBC}} &= A + 2 [(\cos k_y - m) \cos\phi - i \gamma \sin\phi], \\
    \ve^2_{\text{OBC}} &= A + 2 \cos\phi \sqrt{(\cos k_y - m)^2 - \gamma^2} ,
\end{align*}
where $A = 2 + m^2 - \gamma^2 - 2m \cos k_y$. We note that $\ve_{\text{OBC}}$'s are either real or come in complex-conjugate pairs, which can also be traced back to the pseudo-Hermiticity of the real-space Hamiltonian \cite{yao_nh_winding, yao_nh_ci}. The edge spectra given by $\ve_{\L, \R} = \mp \sin k_y$ with purely real decay exponents. Next, the Bloch Hamiltonian [cf. Eq.~(\ref{eq:eigs_bloch})] exhibits second-order Bloch EPs at 
\begin{align}
    & k_x = 0, \quad k_y = \pm \cos^{-1} \left( \frac{(m-1)^2 + 1 - \gamma^2}{2(m-1)} \right),  \label{eq:ci_bloch_ep_one} \\ 
    & k_x = \pi, \quad k_y = \pm \cos^{-1} \left( \frac{(m+1)^2 + 1 - \gamma^2}{2(m+1)} \right).  \label{eq:ci_bloch_ep}
\end{align}
Finally, this system also exhibits a pair of real-space EPs at $k_{y,\text{EP}} = \cos^{-1}(m \mp \gamma)$ by setting $\abs{\Gamma}$ to $0, \infty$, whose order is $N-1$, where $N$ is the system size. At these points, each bulk band collapses to a single point with energy
\begin{equation*}
    \ve = \pm \sqrt{1 + \sin^2 k_{y,\text{EP}}} = \pm \sqrt{2-(m\pm\gamma)^2}.
\end{equation*}
As expected, aside from the qualitative difference, the Bloch and real-space EPs occur at different values of $k_y$ for a given $m$ and $\gamma$. 

We can now analytically deduce the behavior of this system as the non-Hermitian term is turned on. In the following, take $1 < m < 2$, so that we start in a topological phase for $\gamma = 0$. Tuning $\gamma$ up, we nucleate a real-space EP at $k_y = 0$ when $\gamma = m-1$, for which all the states are localized at the left edge. Further increase in $\gamma$ splits this EP into two real-space EPs at $\pm \cos^{-1} (m-\gamma)$, which move out and merge again at $k_y = \pi$ when $\gamma = m+1$. On the other hand, we nucleate a Bloch EP at $k_y = 0$ for $\gamma = 2-m$, which splits into two EPs that merge at $k_y = \pi$ when $\gamma = m$. For a full phase diagram obtained numerically, see Ref.~\cite{ueda_nh_ci}.

In Fig.~\ref{fig:ci_hy_imag}, we plot the PBC and OBC spectrum for the non-Hermitian Chern insulator for a fixed $m$, and we choose two values of $\gamma$ in two different phases: one with only real-space EPs and one with both real-space and Bloch EPs. The spectrum was computed numerically using ED for a finite system with PBC/OBC. We also plot the curves obtained by solving the equations for the bulk spectra for $\phi = 0,\pi$ (blue solid lines) and $\phi = \pm \pi/2$ (blue dashed lines), which follow various contours of the numerically computed bulk bands. We also plot the analytically computed edge spectrum $\ve_{\L,\R}(k_y)$, only a part of which (corresponding to the decay condition on the eigenvalues) are seen for the particular termination used for the OBC calculation. The spectra in Fig.~\ref{fig:ci_hy_imag} for PBC and OBC show vastly different qualitative behavior. For PBC, the system goes from gapped to gapless. This effect can be seen more clearly in a 3D plot of the complex bulk band energies as a function of $k_y$, as shown in the left and middle panel in Fig.~\ref{fig:ci_hy_3d}. The bulk band topology clearly changes from two cylinders, which can be ``flattened'' into two bands, to the \emph{pair of pants}, which cannot be flattened. The OBC spectrum is qualitatively unaffected by this transition. Indeed, we note that the edge states run along a noncontractible loop on the $\ve$-Riemann surface in both cases, as shown in the right panel of Fig.~\ref{fig:ci_hy_3d}. We point out that this behavior was previously observed in Ref.~\cite{ueda_nh_ci}.

The difference between PBC and OBC spectra here can be intuitively understood as the manifestation of a preferred hopping direction, which leads to a pileup of the continuum states at the edges and thereby to an extreme sensitivity to boundary conditions \cite{flore_biorth}. Moreover, when $\gamma$ is chosen such that the hopping in one direction is completely turned off, we get real-space EPs, where the bulk bands indeed collapse to single points, as shown in Fig.~\ref{fig:ci_hy_imag}. The corresponding eigenstates have a finite support, independent of the system size. These EPs are thus associated with an extreme form of unidirectionality.

\subsubsection{Unimodular transfer matrix}
We recall that unimodularity of the transfer matrix implies identical qualitative behavior for the spectrum for PBC and OBC (cf. Sec.~\ref{sec:r1_gamma1}), and thus we only plot band spectra for the latter in this section without loss of information. Moreover, as in this case no real-space EPs may appear, we may make use of the eigenvalues of the Bloch Hamiltonian in Eq.~(\ref{eq:eigs_bloch}) to determine the location of EPs in the spectrum of OBC. 

\begin{figure}[t]
  \includegraphics[width=\columnwidth]{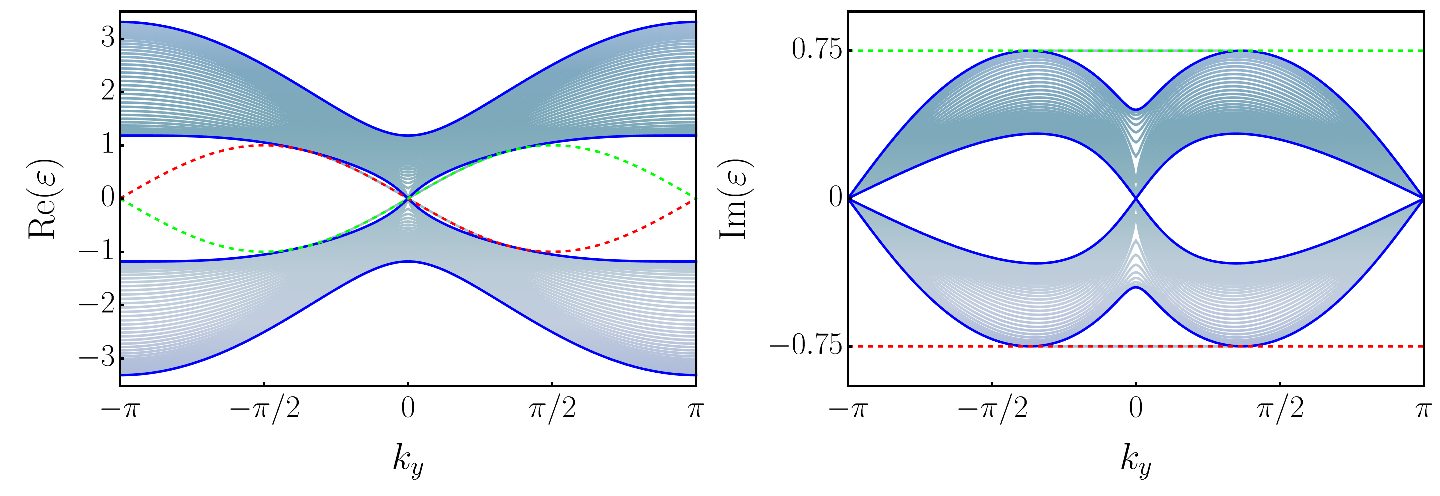}
  \caption{Analytically and numerically computed real and imaginary band structures for the Chern insulator with $N=40$, $m = 1.4$, and $h_z = 0.75$ for OBC. The spectrum for PBC is identical to that for OBC, except for the edge states. We also get a Bloch EP for $k_y = 0$ with both PBC and OBC. 
  }
  \label{fig:ci_hz_imag}

  \ \\ \ \\ 
  \includegraphics[width=\columnwidth]{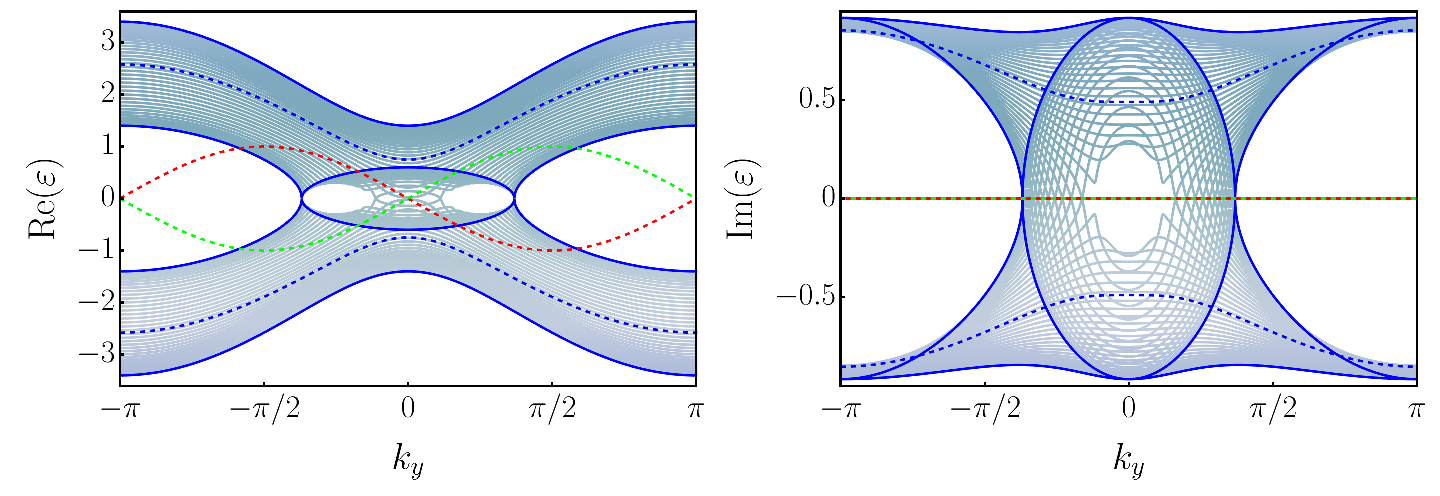} 
  \caption{Analytically and numerically computed real and imaginary band structures for the Chern insulator with $N=40$, $m = 1.4$, and $h_x = \sqrt{0.84}$ for OBC. The spectrum for PBC is identical to that for OBC, except for the edge states.  
  }
  \label{fig:ci_hx_imag}
\end{figure}

We first set $h_z = \gamma$. With a rotation of $\bsg$, the corresponding Bloch Hamiltonian is equivalent to the Bloch Hamiltonian in the case of $h_y = \gamma$ with $k_x$ and $k_y$ interchanged. These two models are thus equivalent up to a $\pi/2$-rotation from a Bloch Hamiltonian perspective, and the EPs are given by expressions identical to the case of $h_y = \gamma$ [cf. Eqs.~(\ref{eq:ci_bloch_ep_one}) and (\ref{eq:ci_bloch_ep})] with the roles of $k_x$ and $k_y$ interchanged. In particular, we find the same behavior for the EPs as we tune $\gamma$, while the systems look completely different from a real-space perspective. The edge spectra are $\ve_{\L, \R} = \mp \left(\sin k_y + i\gamma \right)$, and have now picked up an imaginary part, so that the edge modes now have a finite lifetime. The opposite sign of the imaginary part in the energy of these states is explained by the fact that they are primarily localized on alternate sublattices. Their decay exponents, however, stay real. In Fig.~\ref{fig:ci_hz_imag}, we plot the spectra for OBC with the same parameters as for the previous case, i.e., $m = 1.4$ and $h_z = 0.75$. We note that the EPs appear at $k_y = 0$, which is indeed suggested by Eqs.~(\ref{eq:ci_bloch_ep_one}) and (\ref{eq:ci_bloch_ep}), once the roles of $k_x$ and $k_y$ are interchanged.

We finally set $h_x = \gamma$, so that our model is the usual lattice Dirac equation with a complex mass. The Bloch Hamiltonian, and hence the bulk spectra for both PBC and OBC, exhibit second-order EPs when
\begin{align*}
    & k_x = 0, \quad k_y = \pm \cos^{-1} \left( \frac{2 + (m-i \gamma)^2 - 2(m-i \gamma)}{2(m - i \gamma -1)} \right),  \\ 
    & k_x = \pi, \quad k_y = \pm \cos^{-1} \left( \frac{2 + (m-i \gamma)^2 + 2(m-i \gamma)}{2(m - i \gamma +1)} \right).
\end{align*}
The edge spectra $\ve_{\L, \R} = \mp \sin k_y$ is real, but the corresponding decay exponents now pick up an imaginary part. We find Bloch EPs when the imaginary part of the above equations disappears and the real part is confined to $[-1,1]$.
In Fig.~\ref{fig:ci_hx_imag}, we plot the spectra for OBC with appropriate parameter values and we indeed find Bloch EPs at these values of $k_y$.

\begin{figure}[ht!]
  \centering
    \includegraphics[width=0.5\columnwidth]{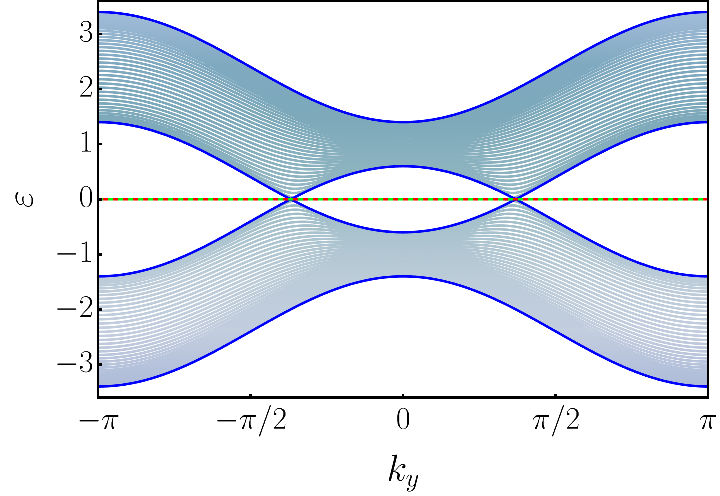} $\quad$
    \includegraphics[width=0.4\columnwidth]{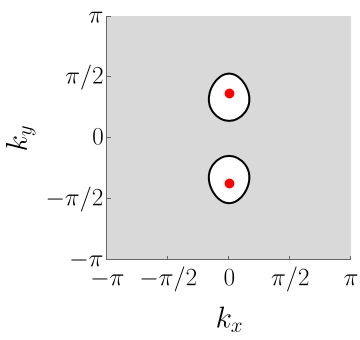} \\ 
    {\includegraphics[width=\columnwidth]{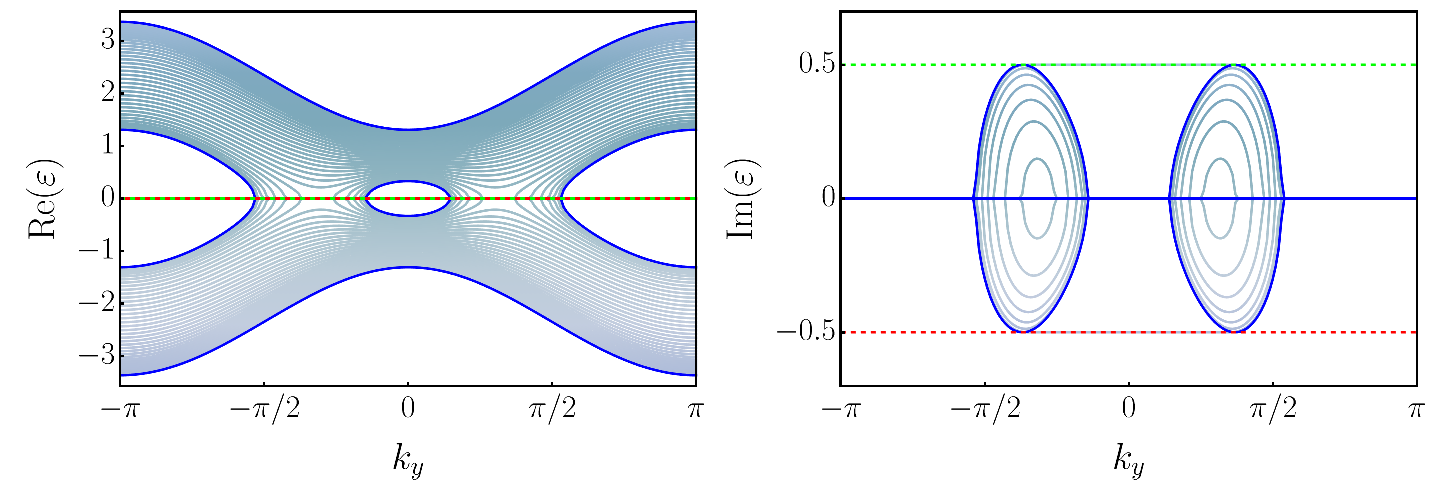}}
  \caption{Analytically and numerically computed band structures for the Dirac semimetal with $N=40$, $m=1.4$, and $\gamma = 0$ (top left) and $\gamma = 0.5$ (bottom). For the latter, the bulk spectrum for PBC is identical to that for OBC. In the top right panel, we show the phase diagram for this model computed from the Bloch spectrum, where the system is in the $\PT$-(un)broken phase in the (gray) white region. The red dots denote the Dirac points for the Hermitian case, which broaden into the exceptional lines denoted by the black solid line as the non-Hermitian term is turned on.}
  \label{fig:pt_model}
\end{figure}

\subsection{A non-Hermitian 2D Dirac semimetal}

In this section, we consider a non-Hermitian lattice model with $\PT$ symmetry, \viz., a two-dimensional Dirac semimetal. This is essentially a two-dimensional stacking of the $\PT$-symmetric Su-Schrieffer-Heeger chains studied in Refs.~\cite{schomerus_pt_ssh, lieu_topological_phases}. Explicitly, we consider the model of Sec.~\ref{sec:r1_model} with
\begin{equation*}
    \bet (k_y) = (\cos k_y - m, \, 0 , \, i \gamma)
\end{equation*}
with $m, \gamma \in \real$. The $\PT$ operation is implemented by $\PTop = \sigma^x K$. Physically, the non-Hermitian term $i\gamma \sigma_z$ in the Bloch Hamiltonian can be understood as a gain on one of the site types and a loss on the other type.

For $\gamma = 0$, we recover the Hermitian limit. In this case, the model is gapped and trivial if $\abs{m} > 2$, while for $\abs{m} < 2$ we get two Dirac points in the 2D Brillouin zone at $\vk = \left( 0, \pm \cos^{-1} (m-1) \right)$ for $0<m<2$ and $\vk = \left( \pi, \pm \cos^{-1} (m+1) \right)$ for $-2<m<0$, which is indeed shown in the top left panel of Fig.~\ref{fig:pt_model}. Turning on the non-Hermitian term $\gamma \neq 0$, these Dirac points broaden into curves of EPs (or \emph{exceptional lines}). Using the eigenvalues of the Bloch Hamiltonian in Eq.~(\ref{eq:eigs_bloch}), we compute that these lie along 
\begin{align}
    (\cos k_x + \cos k_y - m )^2 + \sin^2 k_x - \gamma^2 = 0 \nonumber
\end{align}
These lines separate the $\PT$-unbroken and $\PT$-broken phases as is shown in the phase diagram in the top right panel of Fig.~\ref{fig:pt_model} for $m=1.4$ and $\gamma=0.5$. Explicitly, we have a $\PT$-unbroken phase, i.e., real energies, if the left-hand side is positive, and $\PT$ is broken otherwise. 

From the transfer-matrix perspective, we find  
\begin{equation*}
    \Delta (k_y) = \ve^2 (k_y) - ( \cos k_y - m)^2 + \gamma^2 - 1
\end{equation*}
and $\Gamma = 1$. The latter implies that the bulk spectra for PBC and OBC are identical in both $\PT$-unbroken and $\PT$-broken phases
\footnote{
    Note that in general, this is ensured by the $\PT$-symmetry only in the $\PT$-unbroken phase.
}. 
The bulk spectrum for both PBC and OBC is given by $\Delta = 2\cos\phi$, i.e., the Bloch spectrum. The edge states satisfy $\ve_{\L,\R} = \mp i\gamma$, so that we get a gain for the left edge state and loss for the right one. This is expected, since each of the edge states obtained above is primarily localized on one of these two types of sites. We plot the spectra for OBC in the bottom row of Fig.~\ref{fig:pt_model} with $m=1.4$ and $\gamma =0.5$. We find Bloch EPs for four different values of $k_y$ as predicted by the phase diagram (cf. top right panel of Fig.~\ref{fig:pt_model}]).

\subsection{A non-Hermitian Hofstadter model}

\begin{figure}[t]
  {\includegraphics[width=\columnwidth]{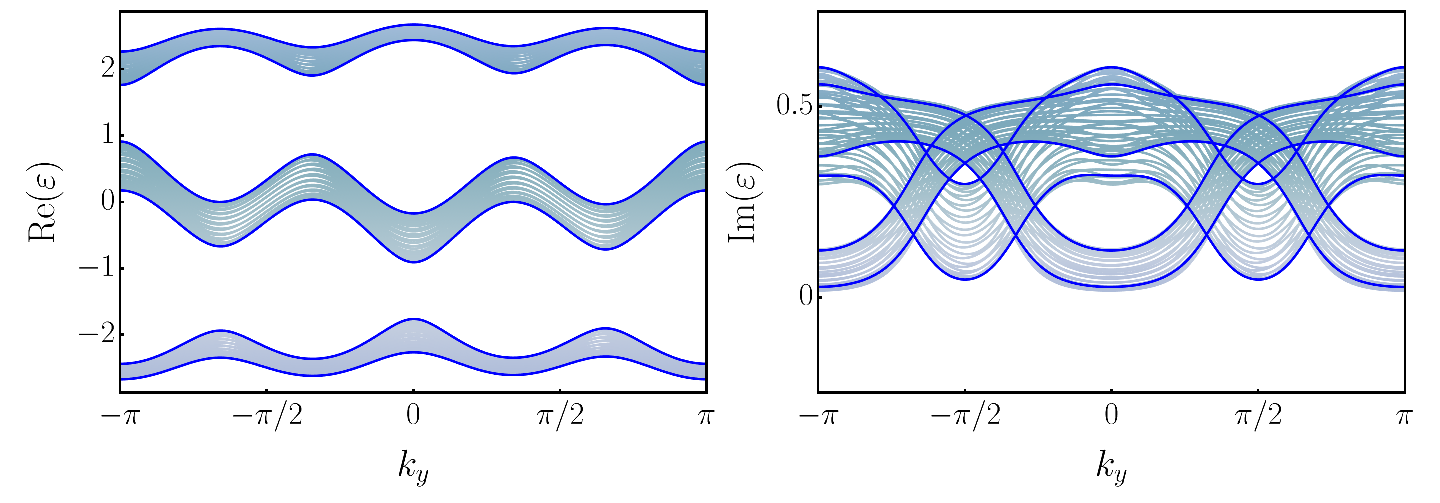}} \\
  {\includegraphics[width=\columnwidth]{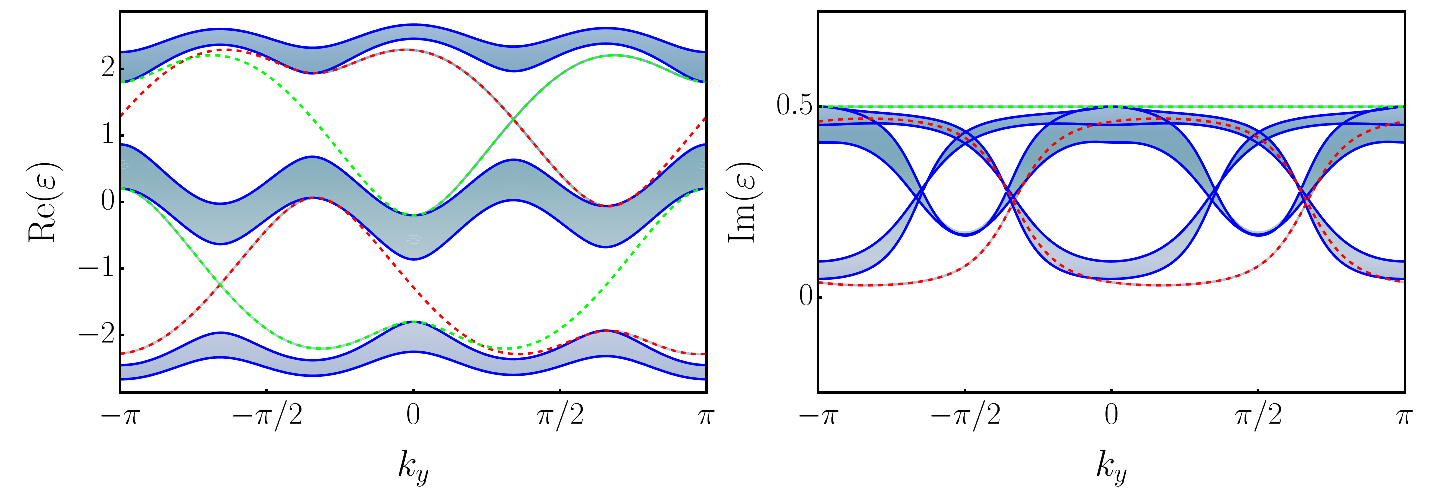}} 
  \caption{The real and imaginary parts of the spectrum for the $\phi = 1/3$ non-Hermitian Hofstadter model on $N=25$ unit cells with PBC (top) and OBC (bottom) for parameters $\gamma_1 = \gamma_2 = 0.5$, $\gamma_3 = 0$, $\delta_1 = 0.6$, and $\delta_2 = -0.25$}
\label{fig:hofstadter}
\end{figure}

We finally consider a non-Hermitian generalization of the Hofstadter model \cite{hatsugai_cbs}, variations of which have also been studied in Refs.~\cite{matveenko_the_area, chernodub_fractal_energy}. Recall that the Hofstadter model is essentially a hopping model on a square lattice with hopping of equal magnitude across each link and phases corresponding to a rational flux of $2\pi \phi$ with $\phi = p/q$ threading each plaquette. We introduce non-Hermiticity in this model either by adding on-site terms $i\gamma_n$ corresponding to absorption/decay and by staggering the magnitude of left hopping vs right hopping by $\delta_n$. Explicitly, assuming translation invariance and PBC along $y$, we consider the Hamiltonian
\begin{align}
    \hlt &= -\sum_n \left[ (1 + \delta_n) \cd_n \c_{n+1} +  (1 - \delta_n) \cd_{n+1} \c_n \right. \nonumber \\ 
    & \left. \quad + \left( 2 \cos (k_y - 2 \pi n \phi) + i\gamma_n^\pdg \right) \cd_n \c_n \right],
\end{align}
where $\gamma_n, \delta_n \in \real$. The original Hofstadter model is periodic with period $q$. To recover this periodicity as well as $J_\L = J_\R$ required for our transfer matrix construction, we choose $\gamma_n = \gamma_{n \, (\text{mod} \,q)}$, $\delta_n = \delta_{n \, (\text{mod} \,q)}$ and $\delta_q = 0$, while the remaining $(2q-1)$ parameters are arbitrary. We can now write the hopping and on-site matrices explicitly. The hopping matrix $J$ has all entries equal to zero except $J_{1,q} = 1$ and satisfies $J^2 = 0$ for all $q>1$. On the other hand, $M$ has $2 t_n \cos (k_y - 2 \pi n \phi) + i\gamma_n$ as its diagonal entries and $(1 \pm \delta_n)$'s on the first diagonal, with $\gamma_n, \delta_n \in \real$. Explicitly, for the simplest nontrivial case of $\phi = 1/3$, we set 
\begin{widetext}
\begin{equation}
    J = 
    \begin{pmatrix}
        0 & 0 & 1 \\    
        0 & 0 & 0 \\   
        0 & 0 & 0
    \end{pmatrix}, \qquad 
    M = 
    \begin{pmatrix}
    2 \cos \left(k_y-\frac{2\pi }{3}\right) + i\gamma_1 & 1+\delta_1 & 0 \\  
    1-\delta_1 & 2 \cos \left(k_y+\frac{2\pi }{3}\right) + i\gamma_2 & 1+\delta_2 \\  
    0 & 1-\delta_2 &  2 \cos (k_y) + i\gamma_3
    \end{pmatrix},
\end{equation}
\end{widetext}
and we find
\begin{equation*}
    \Gamma = \frac{(1+\delta_1) (1+\delta_2)}{(1-\delta_1) (1-\delta_2)}.
\end{equation*}
We plot the spectrum with OBC and PBC in Fig.~\ref{fig:hofstadter}. When we choose $\delta_1 = \pm 1$ (or $\delta_2 = \pm 1$), we find $\Gamma = 0$ or $\infty$ and the continuum bands in the spectrum for the OBC shrink to exceptional lines of order $(N-1)$.

\section{Discussion}
\label{sec:conc}

In this paper, we construct a generalized transfer matrix for non-Hermitian noninteracting tight-binding models and show that various peculiarities of non-Hermitian models are related to simple and readily computable features of the transfer matrix. For instance, the unimodularity of the transfer matrix, a property of Hermitian systems as well as $\PT$-symmetric systems in the $\PT$-unbroken phase, is shown to be related to a bulk-boundary correspondence, while a departure from unimodularity is related to a difference between the PBC and OBC spectra as well as the non-Hermitian skin effect, thereby establishing a formal connection between these two phenomena. These results are illustrated through various examples, which are analytically tractable and highlight the power of this method. For a particular class of systems where the transfer matrix is two-dimensional, we find that the singularity of the transfer matrix is accompanied by the appearance of real-space EPs in the OBC spectrum at which all states are confined to the boundary.

We further find that the topological invariants proposed in Refs.~\cite{yao_nh_winding, yao_nh_winding, yao_nh_ci}, which make use of a deformation of the Brillouin zone to complex quasimomentum, can be naturally understood and generalized in the language of transfer matrices. Explicitly, for $2 \times 2$ transfer matrices, they are obtained straightforwardly by replacing $\e^{ik}$ in the Bloch Hamiltonian with $\sqrt{\det T} \e^{ik}$, i.e., by replacing the eigenvalue of the transfer matrix for PBC with that for OBC. Moreover, the real-space invariant proposed in Ref.~\cite{flore_biorth} in the form of the biorthogonal polarization can also be readily obtained making use of transfer matrices. Additionally, we show that, at least for two-dimensional systems, we can assign a topological invariant to the edge states by identifying the edge spectra $\ve_{\L,\R}(k_y)$ as closed loops on the energy Riemann surface. If these loops are noncontractible, the edge modes can only be removed if the bulk gap for OBC collapses, which may be independent of the bulk gap for PBC for non-Hermitian systems. This is indeed the case for the non-Hermitian Chern insulator model studied in Sec.~\ref{sec:eg_ci_hy}, where a gap closing in the PBC spectrum leaves the edge state unaffected. Therefore, the transfer matrices, which give access to the eigensystem of the model with open boundary conditions, provide a crucial insight to establish the topology of these models.

Interestingly, the extension to non-Hermitian Hamiltonians of many of the results previously obtained in Ref.~\cite{vd-vc_tm} for Hermitian models involves several aspects that were not needed or not quite visible in the Hermitian case. For instance, the bulk spectra for PBC and OBC correspond to very different mathematical conditions, which happen to coincide when the transfer matrix is unimodular. A further attraction of this generalization is the possibility of complex energies, which lends a physical significance to the construction of an energy Riemann surface, which was introduced for Hermitian systems purely for mathematical convenience. 

The explicit computations for the rank-1 case discussed in Sec.~\ref{sec:r1} do not readily generalize to higher ranks. In certain cases, however, additional structure in the model can be leveraged. For instance, if the transfer matrix happens to be symplectic, then the spectra can be obtained by using the restriction on the eigenvalue problem that it imposes \cite[Appendix C]{vd-vc_tm}. In absence of such additional structure, the spectra can be computed directly using the results of Sec.~\ref{sec:tm_spec}, in particular, Eqs.~(\ref{eq:pbc1}) and (\ref{eq:obc_cramers}). Both of these require the diagonalization of the transfer matrix and hence should be numerically, if not analytically, tractable. A more algebraic approach to these cases, especially the construction of associated Riemann surfaces and definition of boundary invariants, requires further study. 

We emphasize that the transfer-matrix approach is also useful for systems not described by tight-binding models. For instance, transfer matrices have been extensively used to study localization in the phenomenological Chalker-Coddington network models \cite{evers2008anderson}. A non-Hermitian version thereof was also studied in Ref.~\cite{brouwer_theory_of} for a one-dimensional periodic chain with an imaginary vector potential. We believe that the insights gleaned from our extension of transfer-matrix formalism to non-Hermitian systems would prove useful in diverse contexts.

One particularly interesting direction for further investigation is the implications of symmetries on the transfer matrix. Indeed, one of the central parts of the study of Hermitian topological phases has focused on their classification, e.g, gapped non-interacting fermionic Hamiltonians belong to one of the 10 equivalence (Altland-Zirnbauer) classes based on their antiunitary symmetries \cite{altland-zirnbauer, ryu-furusaki}. The non-Hermitian analogs of this ``ten-fold way'' are described by Bernard and LeClair \cite{bernard-leclair, *bernard-leclair_book}, and allow for many more symmetries. Indeed, recent investigations of non-Hermitian Bloch Hamiltonians have resulted in their classification in terms of genuinely non-Hermitian symmetry classes \cite{ueda_nh_ktheory, kawabata_nh_tsc}, while in previous studies explicit non-Hermitian topological phases with a trivial Hermitian limit have already been constructed \cite{flore_symmetry_protected}, and the role of time-reversal symmetry was also investigated in Ref.~\cite{sato_edge_states}. The transfer-matrix approach can shed further light on the general classification of non-Hermitian Hamiltonians, since as we show in this paper for $\PT$ symmetry, the symmetries of the Hamiltonian may be implemented in a nontrivial manner on the transfer matrix. Moreover, the possible difference between spectra obtained for PBC and OBC makes a classification of systems based on a real-space approach highly relevant.

The discrepancy between periodic and open spectra necessitates the access to exact solutions of the real-space Hamiltonian in order to probe the topological aspects of a system. The transfer-matrix approach, being a purely real-space construction, is the ideal platform for such an endeavor. We thus believe that transfer matrices provide a natural framework for a general understanding of non-Hermitian systems.

\acknowledgments{
We thank Emil J. Bergholtz, Victor Chua, Henry Legg, Max Geier, Sebastian Diehl, and Thors Hans Hansson for useful conversations.
FKK was funded by the Swedish Research Council (VR) and the Knut and Alice Wallenberg Foundation.
VD was funded by the \textit{Deutsche Forschungsgemeinschaft} (DFG, German Research Foundation) -- Projektnummer 277101999 -- TRR 183 (project B03).
}

\appendix 

\section{Non-Hermitian Hamiltonians and \texorpdfstring{$\PT$}{} symmetry} \label{app:pt_symm}

Non-Hermitian systems are described by Hamiltonians with $\hlt \neq \hlt^\dg$, with the adjoint taken under the usual inner product on the Hilbert space. Consequently, the left and right eigenstates, defined by 
\[ 
    \hlt \ket{\psi_n} = \ve_n \ket{\psi_n}, \qquad \bra{\phi_n} \hlt = \ve_n \bra{\phi_n},
\]
are not related by conjugation and $\braket{\psi_m | \psi_n} \neq \delta_{mn}$, i.e., the eigenbasis is no longer orthonormal. However, $\braket{\phi_m | \psi_n} = \delta_{mn}$, so that one defines a biorthogonal basis and uses it to compute the so-called \emph{biorthogonal} expectation value \cite{brody_biorthogonal_quantum} of an observable $\mathcal{O}$ as $\braket{\phi | \mathcal{O} | \psi}$. The non-orthogonality of eigenvectors may also lead to two or more eigenvectors becoming linearly dependent, so that the eigenstates do not span the Hilbert space. Such Hamiltonians are termed \emph{defective}. For the Hamiltonian dependent on a set of parameters, their value for which the Hamiltonian is defective is termed an \emph{exceptional point} (EP) \cite{ kato_book, heiss_EPs}, whose order is defined as the number of eigenvectors that coalesce at that EP.

Systems with a parity-time-reversal ($\PT$) symmetry \cite{mathur_pt_symm} form a particularly well-studied subset of non-Hermitian systems. 
$\PT$ is implemented as an antilinear and antiunitary operator $\PTop = \uop K$ with $\uop$ a unitary matrix and $K$ denoting complex conjugation. Furthermore, $\left( \PTop \right)^2 = \pm \id$, which corresponds to $\uop^T = \pm \, \uop$. Owing to its antilinearity, $\PTop$ cannot have eigenvectors. Explicitly, this is because if $\psi$ were an eigenvector, so would $a \psi$ for any $a \in \cmplx$, but
\[
    \PTop\psi = \uop K \psi = \rho \psi \implies \PTop a \psi = a^\ast \rho\psi.
\]
However, a state may be left invariant up to a phase under $\PT$. A further constraint is imposed by 
\[
     (\PTop)^2\psi = \PTop \, \rho \psi = \abs{\rho}^2 \psi,
\]
so that $\left(\PTop\right)^2 = -\id$, there are no states left invariant under $\PTop$, while for $\left(\PTop\right)^2 = \id$, one might have such states. 

Under the $\PT$ operation, the Bloch Hamiltonian transforms as 
\begin{equation}
    \PTop\colon \hlt_B(\vk) \mapsto \uop \,  \hlt_B^\ast(\vk) \, \uop^\dg. 
    \label{eq:pt_action}
\end{equation}
If $\hlt_\B$ is $\PT$-symmetric, the states must satisfy 
\begin{equation}
    \hlt_B \psi = \ve \psi \iff 
    \hlt_B \, \PTop \psi = \ve^\ast \, \PTop \psi,
    \label{eq:pt_cond_wf}
\end{equation}
where $\ve \in \cmplx$ in general. For $\left(\PTop\right)^2 = \id$, a state may be invariant under $\PT$. In this case, the two eigenvalue equations in Eq.~(\ref{eq:pt_cond_wf}) contain the same eigenvector, so that $\ve = \ve^\ast$. When this is true for all eigenstates, the system is termed to be in a \emph{$\PT$-unbroken phase}. On the other hand, if there are eigenstates that are not invariant under $\PT$, i.e., the $\PT$ symmetry is spontaneously broken, then we are in a \emph{$\PT$-broken} phase. For $\left(\PTop\right)^2 = -\id$, no state is left invariant by $\PT$, so that the system is always in a $\PT$-broken phase. The states satisfy an analog of Kramers' theorem \cite{mathur_pt_symm}, \viz., all eigenstates come in orthogonal pairs whose energies are related by complex conjugation. A trivial consequence of this case is that the Bloch Hamiltonian must be even dimensional.   

\section{\texorpdfstring{$\PT$}{PT}-symmetry and SVD}  
\label{app:svd}

We derive the constraints imposed by the $\PT$-symmetry on the singular values and vectors of the hopping matrix. Recall that the reduced SVD of a matrix $J \in \Mat(\nuc, \cmplx)$ is defined as \cite[Sec.~6.3]{strang_book}
\begin{equation}
    J = V \, \Xi \, W^\dg
    = \sum_{n=1}^r \xi_n \vv_n \vw_n^\dg,
\end{equation}
where $r = \rank J$, $\xi_n > 0$ are the singular values and $\vv_i, \vw_i$ the corresponding left/right singular vectors. A pair of singular vectors $\vv, \vw$ is defined by the relations
\begin{equation}
    J \vw = \xi \vv, \qquad 
    J^\dg \vv = \xi \vw.
    \label{eq:svec_def}
\end{equation}
Note that these expressions are manifestly invariant under a simultaneous phase rotation $\vv \to \e^{i\theta} \vv, \; \vw \to \e^{i\theta} \vw$. 

In presence of a $\PT$ symmetry, the hopping matrix satisfies $J^T = \uop^\dg J \uop$ with $\uop \in \U(\nuc)$, so that Eq.~(\ref{eq:svec_def}) becomes
\begin{align*}
J^\ast \uop^\dg \vv &= \uop^\dg J^\dg \vv = \xi \uop^\dg \vw, \\
    J^T \uop^\dg \vw &= \uop^\dg J \vw = \xi \uop^\dg \vv. 
\end{align*}
A complex conjugation leads to
\[
    J \wt{\vw} = \xi \wt{\vv}, \qquad 
    J^\dg \wt{\vv} = \xi \wt{\vw},
\]
where $\wt{\vw} = \uop^T \vv^\ast$ and $\wt{\vv} = \uop^T \vw^\ast$. We thus find two sets of vectors satisfying the equation for a singular value $\xi$, so that either the two vectors are proportional, i.e., 
\[
    \exists \, \rho \in \cmplx \;\; \text{such that} \;\;  \vv = \lambda \wt{\vv} \iff   \vw = \lambda \wt{\vw},
\]
or $\xi$ is degenerate as a singular value with two sets of left and right singular vectors. When the two vectors are proportional, we find 
\begin{equation}
    \vv = \lambda \, \uop^T \vw^\ast 
    = \abs{\lambda}^2 \uop^T \uop^\dg \vv. 
    \label{eq:l_cond}
\end{equation}
Here, we consider the two possible cases: If $\uop^T = \uop$, then Eq.~(\ref{eq:l_cond}) holds iff $\abs{\lambda} = 1$. Setting $\lambda = \e^{2i\chi}$, we get $\vv = \e^{2i\chi} \uop^T \vw^\ast$, and we can use the invariance of \eq{eq:svec_def} under the phase rotation $\vv \to \vv \e^{i\chi}, \, \vw \to \vw \e^{i\chi}$ to fix the phase of $\vv, \vw$ such that $\vv = \uop \vw^\ast$ and $\vw = \uop \vv^\ast$. Continuing this for all singular vectors of $J$, we find 
\[
    V = \uop \, W^\ast, \qquad 
    W = \uop \, V^\ast, 
\]
which are the requisite conditions on the singular vectors of $J$. 

On the other hand, if $\uop^T = -\uop$, then $\nexists \, \lambda\in\cmplx$ for which Eq.~(\ref{eq:l_cond}) holds. Therefore, the singular value $\xi$ must be degenerate with the corresponding right and left singular vectors reading $\vw, \wt{\vw}$ and $\vv, \wt{\vv}$, respectively. In this degenerate sector, we set $\mathfrak{v} = (\vv, \wt{\vv})$ and $\mathfrak{w} = (\vw, \wt{\vw})$ to write the SVD of $J$ in this subspace as $\mathfrak{v} \, \id_2 \, \mathfrak{w}$. Using the definition of $\wt{\vv}$ and $\wt{\vw}$, this leads to
\[
    \mathfrak{v} 
    = (\uop \wt{\vw}^\ast, -\uop \vw^\ast)
    = \uop \mathfrak{w}^\ast \mathscr{J}, \quad \mathscr{J} = 
    \begin{pmatrix}
        0 & -1 \\ 
        1 & 0 
    \end{pmatrix}.
\]
$J$ thus falls apart into these $2\times 2$ $\mathfrak{v} \, \id_2 \, \mathfrak{w}$ blocks with degenerate singular values, so that $r = \rank J$ must be even. Defining $\Sigma = \mathscr{J} \otimes \id_{r/2}$, we find 
\[
    V = \uop \, W^\ast \, \Sigma, \qquad 
    W = \uop \, V^\ast \, \Sigma, 
\]
which are the requisite conditions on the singular vectors of $J$.

\section{Schur complement and inversion}
\label{app:schur}
Here, we explore some of the algebraic properties of the rank-1 transfer matrix. Given the on-site matrix $M \in \Mat(\nuc, \cmplx)$ and energy $\ve\in\cmplx$, the on-site Green's function can be written as 
\begin{equation}
    \green \equiv (\ve \id - M)^{-1} = \frac{1}{Q(\ve)} G(\ve),
\end{equation}
where $Q(\ve) \equiv \det (\ve \id - M)$ is by definition a polynomial in $\ve$ of order $\nuc$, and $G(\ve) \equiv \text{adj}(\ve \id - M) \in \Mat(\nuc, \cmplx)$ is the \emph{adjugate} (i.e., the matrix of minors) \cite[Sec.~4.4.1]{strang_book} of $M$, whose elements are polynomials in $\ve$ of order $\leq \nuc - 1$. The transfer matrix can be written as 
\begin{equation}
  T = \frac{1}{\xi G_{vw}}
  \begin{pmatrix} 
    Q & \;  - G_{ww} \xi \\ 
     G_{vv} \xi & \; \frac{\xi^2}{Q}\left( G_{vw} G_{wv} - G_{vv} G_{ww} \right),
  \end{pmatrix},
\end{equation}
where $G_{ab}(\ve) \equiv Q(\ve) \, \green_{ab} (\ve) , \; a,b \in \{ \vv, \vw \}$ are polynomials (instead of rational functions) in $\ve$. The discriminant becomes 
\begin{align}
    \Delta^2 - 4\Gamma &= \frac{1}{\xi^2 G_{vw}^2} \bigg[ \left( Q + \frac{\xi^2}{Q}\left( G_{vw} G_{wv} - G_{vv} G_{ww} \right) \right)^2 \nonumber \\ 
    & \qquad\qquad\qquad - 4 \xi^2 \; G_{vw} G_{wv} \bigg],
\end{align}
and we are interested in its zeros. Naively, owing to the $G_{vv}^2 G_{ww}^2$ term, the numerator is a polynomial of order $ \leq \ve^{4(\nuc-1)}$ in $\ve$. However, in the following we show that 
\begin{equation}
    f(\ve) \equiv \frac{1}{Q(\ve)} \left( G_{vw} G_{wv} - G_{vv} G_{ww} \right)
\end{equation}
is a polynomial in $\ve$ of order $ \leq \nuc -1$, so that the leading-order term in the numerator of $\Delta^2 - 4\Gamma$ arises from $Q^2$, rendering it a polynomial in $\ve$ of order $2\nuc$.

We begin by using the basis independence of the transfer-matrix computation to choose a basis of $\cmplx^\nuc$ in which $\vv = (1,0,0, \dots)$ and $\vw = (0,1,0, \dots)$, so that
\begin{equation}
    G(\ve) = Q(\ve) \, \green(\ve) = 
    \begin{pmatrix}
        A & B \\ 
        C & D
    \end{pmatrix},   \label{eq:G_block}
\end{equation}
where $A \in \Mat(2, \cmplx)$ and the numerator of $f(\ve)$ is simply $\det A$. Using the fact that for block matrices
\begin{equation}
     \det G = \det A \, \det S, \quad S \equiv D - C A^{-1} B,
\end{equation}
we can rewrite $f(\ve)$ as  
\begin{equation}
    f(\ve) = \frac{\det A(\ve)}{Q(\ve)} 
    = \frac{\det G(\ve)}{Q(\ve) \det G(\ve)} 
    = \frac{Q^{\nuc - 2}(\ve)}{\det S(\ve)},
\end{equation}
where we have used the fact that $\det G = Q^{\nuc-1}$. Using the inversion formula for block matrices \cite[Eq.~(A8)]{vd-vc_tm}, Eq.~(\ref{eq:G_block}) becomes 
\begin{equation}
    G^{-1} = 
    \begin{pmatrix}
        A^{-1} + A^{-1} B S^{-1} C A^{-1} & -A^{-1} B S^{-1} \\ 
        S^{-1} C A^{-1}  & S^{-1}
    \end{pmatrix}.
\end{equation}
But we also have 
\begin{equation}
    G^{-1}(\ve) = \frac{1}{Q(\ve)} (\ve \id - M).
\end{equation}
We thus identify $S^{-1}$ as the $(\nuc-2)\times(\nuc-2)$ lower right term in the block structure of $(\ve \id - M)$. Finally, 
\begin{equation}
    f(\ve) = Q^{(\nuc-2)}(\ve) \det S^{-1}(\ve) = \det \left( \ve \id - M \right)_X,
\end{equation}
where $\left( . \right)_X$ denotes the restriction to the subspace spanned by the orthogonal complement of $\vv$ and $\vw$. Thus, $f(\ve)$ is a polynomial in $\ve$ of order $\nuc - 2$, which proves our desired result.

\section{Explicit computations for \texorpdfstring{$r=1$}{r=1}}
\label{app:obc}

We compute $T^n (\ve)$ explicitly for $T \in \Mat(2,\cmplx)$ and arbitrary $n \in \intg$ and use it to derive explicit conditions on $\ve$ for obtaining an eigenstate of a system with OBC. 

\subsection{Computing \texorpdfstring{$T^n$}{}}
We start off with Cayley's theorem, which states that a matrix satisfies its characteristic equation. Thus, $T \in \Mat(2,\cmplx)$ satisfies 
\begin{equation}
    T^2 - \Delta \, T + \Gamma \id = 0, 
    \label{eq:tm_char}
\end{equation}
where $\Delta = \tr \, T$ and $\Gamma = \det T$. For $\Gamma = 0$, we simply get $T^n = \Delta^{n-1} T$. On the other hand, for $\Gamma \neq 0$, using Eq.~(\ref{eq:tm_char}) repeatedly, one can reduce $T^n = A_n T + B_n \id$. Using $T^{n+1} = T \, T^n$, i.e., 
\begin{equation*}
    A_{n+1} T + B_{n+1} \id = (A_n \Delta + B_n) T - A_n \Gamma \id,
\end{equation*}
we obtain a recursion relation for the coefficients 
\begin{align*}
    A_{n+1} &= A_n \Delta + B_{n}, \\
    B_{n+1} &= -A_n \Gamma.
\end{align*}
These reduce to a three-term recursion for $A_n$ as
\begin{equation}
    A_{n+1} = A_n \Delta - A_{n-1} \Gamma,
\end{equation}
with the initial condition $A_1 = 1$ and $A_2 = \Delta$. For $\Gamma \neq 0$, setting $A_{n} = \Gamma^{(n-1)/2} a_n$, this reduces to
\begin{equation}
    a_{n+1} = 2z \, a_n - a_{n-1}; \qquad  
    z = \frac{\Delta}{2\sqrt{\Gamma}}
    \label{eq:cheby_recur}
\end{equation}
with the initial conditions $a_1 = 1$ and $a_2 = 2z$. This is the defining relation for the \emph{Chebyshev polynomials of the second kind} $U_n(z)$ \cite[Sec.~10.11]{bateman_spcl}, so that we identify $a_n = U_{n-1}(z)$. This leads to our final result 
\begin{equation}
    T^n = \Gamma^{n/2} \left[\frac{U_{n-1}(z)}{\sqrt{\Gamma}} T - U_{n-2}(z) \id \right], 
    \label{eq:tm_exp}
\end{equation}
which can be easily evaluated using the closed form expressions for the Chebyshev polynomials \cite[Sec.~10.11, Eqn.~(2)]{bateman_spcl}
\begin{equation}
    U_n(z) = \frac{\lambda^{n+1} - \lambda^{-(n+1)}}{2(\lambda-\lambda^{-1})} 
    = \frac{\sin \left( (n+1)\phi \right) }{\sin \phi},
    \label{eq:cheby_def}
\end{equation}
where 
\begin{equation*}
    z = \frac{\lambda + \lambda^{-1}}{2} = \cos\phi.
\end{equation*}
The former expression for $U_n(z)$ in Eq.~(\ref{eq:cheby_def}) is useful for arbitrary $z\in\cmplx$, while the latter is naturally more useful when $z\in\real$ and $\abs{z} < 1$.

\subsection{Open Boundary conditions}
For $\Gamma = 0$, using $T^n = \Delta^{n-1} T$, we trivially get
\begin{equation*}
    \frac{\Delta^{N-1}}{\xi \, \green_{vw}} \vecenv{1}{\xi \green_{vv}} = \mathfrak{r} \vecenv{0}{1}.
\end{equation*}
Thus, the bulk spectrum collapses to a single point given by $\Delta = \Gamma = 0$. For $\Gamma\neq 0$, using the explicit form of $T$ [cf. Eq.~(\ref{eq:rank1_tm})], the condition for OBC in Eq.~(\ref{eq:r1_obc1}) becomes 
\begin{equation}
    \frac{\Gamma^{N/2}}{q} \vecenv{U_{N-1}(z) - q \, U_{N-2}(z) }{\xi \green_{vv}U_{N-1}(z)} = \mathfrak{r} \vecenv{0}{1},   \label{eq:app_obc_cond0}
\end{equation}
where $\mathfrak{r}$ is arbitrary and we have defined
\begin{equation*}
    q = \xi\sqrt{\green_{vw} \green_{wv}}, \quad 
    z = \frac{\Delta}{2\sqrt{\Gamma}} 
    = \frac{1 + q^2 - \xi^2 \green_{vv} \green_{ww}}{2q}.
    \label{eq:app_zdef}
\end{equation*}
The condition on $\ve$ can now be written as 
\begin{equation}
    q = \frac{U_{N-1}(z)}{U_{N-2}(z)}.
    \label{eq:app_obc_cond}
\end{equation}
This can be recast in another useful form by substituting Eq.~(\ref{eq:app_obc_cond}) in Eq.~(\ref{eq:app_zdef}). We get
\begin{align}
    \xi \sqrt{\green_{vv} \green_{ww}} 
    & = \sqrt{q^2 - 2 q z + 1} \nonumber \\ 
    & = \frac{ \sqrt{U_{N-2}^2(z) -  U_{N-1}(z) U_{N-3}(z)}}{U_{N-2}(z)} \nonumber \\
    & = \frac{ \sqrt{ \sum_{k=0}^{N-2} U_{2k}(z) - \sum_{k=0}^{N-3} U_{2k+2}(z) }}{U_{N-2}(z)} \nonumber \\
    & = \frac{ 1}{U_{N-2}(z)},
\end{align}
where we have used the recursion relation for the Chebyshev polynomial in Eq.~(\ref{eq:cheby_recur}) as well as the product formula 
\begin{equation*}
    U_m(z) U_{n}(z) = \sum_{k=0}^n U_{m-n+2k}(z); \quad  m \geq n.
\end{equation*}
In the last step, we have used the fact that $U_0(z) = 1$.

The conditions for OBC can be further reduced in the large-$N$ limit. Using the first definition of Chebyshev polynomials from Eq.~(\ref{eq:cheby_def}), we have
\begin{equation}
    q = \frac{\lambda^{N} - \lambda^{-N}}{\lambda^{N-1} - \lambda^{-(N-1)}}.
\end{equation}
For $N\to\infty$, we need to consider three cases. For $\abs{\lambda} > 1$, we can compute 
\begin{equation*}
    q = \lim_{N\to\infty} \lambda \, \frac{1 - \lambda^{-2N}}{1 - \lambda^{-2(N-1)}} = \lambda = z + \sqrt{z^2-1},
\end{equation*}
while for $\abs{\lambda} < 1$, we get 
\begin{equation*}
    q = \lim_{N\to\infty} \frac{1}{\lambda} \, \frac{\lambda^{2N} - 1}{\lambda^{2(N-1)} - 1} = \frac{1}{\lambda} = z - \sqrt{z^2-1}.
\end{equation*}
Finally, for $\abs{\lambda} = 1$, setting $\lambda = \e^{i\phi}$, we get
\begin{equation*}
    q = \frac{\e^{iN\phi} - \e^{-iN\phi}}{\e^{i(N-1)\phi} - \e^{-i(N-1)\phi}} 
    = \frac{\sin \left( N\phi \right) }{\sin \left( (N-1)\phi \right)},
\end{equation*}
which does not have a well-defined limit as $N\to\infty$; instead, the right hand side oscillates wildly, since it has zeros at $\phi = k\pi/N$ and poles at $k\pi/(N-1)$. Thus, for any $q(\phi)$, we get $N$ solutions in $\phi\in[0,\pi)$, which become dense in $[0,2\pi)$ as $N\to\infty$. This is our bulk band for OBC. In terms of $z$, this also corresponds to setting $z = \cos\phi$. 

We can also derive a condition for a boundary condition interpolating between PBC and OBC [cf. Sec~\ref{sec:tm_spec}], for we demand that $1 \in \spec{K T^N}$. Since $KT^N$ is a $2 \times 2$ matrix, its two eigenvalues must be $1$ and $\det(KT^N) = \det K \left(\det T\right)^N = \Gamma^N$, so that 
\begin{equation}
    \tr \left( KT^N \right) = 1 + \Gamma^N = 2 \, \Gamma^{N/2} \cosh (N\zeta_0),  \label{eq:kbrdy_cond}
\end{equation}
where $\zeta_0 = \frac{1}{2} \log \Gamma$. We next compute the left hand side using Eq.~(\ref{eq:tm_exp}) as 
\begin{align*}
    \tr \left( KT^N \right) 
    &= \Gamma^{N/2} \left[\frac{U_{N-1}(z)}{\sqrt{\Gamma}} \tr(KT) - U_{N-2}(z) \tr(K) \right] \\ 
    &= \Gamma^{N/2} \left[U_{N-1}(z) \left( \frac{1}{q\kappa} + \frac{\kappa}{q} (2qz-1) \right) \right. \\ 
    & \left. \qquad\qquad - U_{N-2}(z) \left( \frac{1}{\kappa} + \kappa \right) \right],
\end{align*}
where we have used 
\begin{equation*}
    \xi^2 \left( \green_{vw} \green_{wv} - \green_{vv} \green_{vw} \right) = \frac{q \, \Delta}{\sqrt{\Gamma}} - 1 = 2 q z - 1. 
\end{equation*}
Using the recursion relation for the Chebyshev polynomials to replace  
\[
    2z \, U_{N-1}(z) = U_{N}(z) + U_{N-2}(z),
\]
Eq.~(\ref{eq:kbrdy_cond}) can be reduced to 
\begin{align*}
    & \cosh (N\zeta_0)  \\ 
    & \qquad = \kappa \, U_N(z) -  \frac{1}{q} \left( \kappa - \frac{1}{\kappa} \right) U_{N-1}(z) - \frac{1}{\kappa} \, U_{N-2}(z).
\end{align*}
Setting $z = \cos \chi$ for some $\chi = \phi + i\zeta; \, \phi, \zeta \in \real$ and using the definition of $U_n(\cos\chi)$, this can be rearranged to get
\begin{align}
    & \left( \kappa - \frac{1}{\kappa} \right) \left( \cot \chi - \frac{\csc\chi}{q} \right) \nonumber \\ 
    & \qquad\qquad = 2 \, \frac{\cosh(N\zeta_0)}{\sin(N\chi)} - \left( \kappa + \frac{1}{\kappa} \right) \cot(N\chi).
\end{align}
The left hand side is now independent of $N$. For the bulk states, we shall require that the right hand side does not have a limit as $N\to\infty$ (as in the OBC case above). Thus, the condition for an eigenstate becomes $\Delta = 2 \sqrt{\Gamma} \cos (\phi + i \zeta)$ for some $\phi \in [0,\pi]$, with
\begin{equation}
    \zeta \approx \zeta_0 - \frac{2}{N} \log \left( \frac{\kappa + \kappa^{-1}}{2} \right)
\end{equation}
for $\kappa$ close to 1. Therefore, the spectrum for $\kappa \neq 0,1$, unlike for the PBC and OBC case, is in general quite sensitive to the system size.

\bibliography{nonhermitian}

\end{document}